\documentclass{sn-jnl}
\usepackage[authoryear,round]{natbib}

\usepackage{graphicx}%
\usepackage{multirow}%
\usepackage{amsmath,amssymb,amsfonts}%
\usepackage{amsthm}%
\usepackage{mathrsfs}%
\usepackage[title]{appendix}%
\usepackage{xcolor}%
\usepackage{textcomp}%
\usepackage{manyfoot}%
\usepackage{booktabs}%
\usepackage{algorithm}%
\usepackage{algorithmicx}%
\usepackage{algpseudocode}%
\usepackage{listings}%
\usepackage{geometry}%
\usepackage{overpic}%

\begin{document}
% \jyear{2024}%

\title[Article Title]{Methodological Reconstruction of Historical Landslide Tsunamis Using Bayesian Inference}

\author[1]{\fnm{Raelynn} \sur{Wonnacott}}\email{raelynnwo@gmail.com}

\author[2]{\fnm{Dallin} \sur{Stewart}}\email{dallinpstewart@gmail.com}

\author*[2]{\fnm{Jared P.} \sur{Whitehead}}\email{whitehead@mathematics.byu.edu}
\author[3]{\fnm{Ronald A.} \sur{Harris}}\email{rharris@byu.edu}

\affil[1]{\orgdiv{Mathematics}, \orgname{University of Maryland}, \orgaddress{\street{4176 Campus Dr.}, \city{College Park}, \postcode{20742}, \state{Maryland}, \country{United States}}}

\affil[2]{\orgdiv{Mathematics}, \orgname{Brigham Young University}, \orgaddress{\street{275 TMCB}, \city{Provo}, \postcode{84602}, \state{Utah}, \country{United States}}}

\affil[3]{\orgdiv{Geology}, \orgname{Brigham Young University}, \orgaddress{\street{
S-349 ESC}, \city{Provo}, \postcode{84602}, \state{Utah}, \country{United States}}}

\abstract{
Indonesia is one of the world's most densely populated regions and lies among the epicenters of Earth's greatest natural hazards.
Effectively reducing the disaster potential of these hazards through resource allocation and preparedness first requires an analysis of the risk factors of the region. 
Since destructive tsunamis present one of the most eminent dangers to coastal communities, understanding their sources and geological history is necessary to determine the potential future risk. %This preparation can inform strategies for implementing measure to reduce risks and save lives.

Inspired by results from Cummins et al. \citep{Cummins2020}, and previous efforts that identified source parameters for earthquake-generated tsunamis, we consider landslide-generated tsunamis. 
This is done by constructing a probability distribution of potential landslide sources based on anecdotal observations of the 1852 Banda Sea tsunami, using Bayesian inference and scientific computing.
After collecting over 100,000 samples (simulating 100,000 landslide induced tsunamis), we conclude that a landslide event provides a reasonable match to the tsunami reported in the anecdotal accounts. However, the most viable landslides may push the boundaries of geological plausibility. Future work creating a joint landslide-earthquake model may compensate for the weaknesses associated with an individual landslide or earthquake source event.}
\label{sec:abstract}

\label{sec:keywords}

\maketitle

\section{Introduction}\label{sec:intro}
On Friday, Sept. 28, 2018, a 7.5 Mw earthquake hit Central Sulawesi in the Indonesian archipelago. The resultant tsunami in Palu bay grew significantly larger than anticipated purely from the size and nature of the earthquake \citep{muhari2018solving}. Several additional studies (see \cite{takagi2019analysis,liu2020coastal,aranguiz20202018,pranantyo2021complex,schambach2021new} for just a few examples) have investigated the probability that this anomaly was due to a combination of the source earthquake and consequent submarine landslides (often referred to as submarine mass failures in the literature). Such a recent, destructive seismic event highlights the need to better understand the potential for submarine landslides to generate dangerous tsunamis in coastal regions. In particular, it is imperative to re-evaluate past records and accounts of tsunamis to determine if those events were due to the seismic activity alone, or if a landslide contributed to the development of the wave itself.

Motivated by the findings from \cite{Cummins2020} we explore the possibility that a submarine landslide caused the 1852 Banda Sea tsunami which Dutch colonists recorded extensively in settlements throughout the region. In doing so, we merge the hypothesis of a submarine landslide with the Bayesian methodology introduced in \cite{ringer2021} and \cite{paskett2023} to identify possible parameters that best model the causal submarine landslide. Our goal in this study is to determine if a submarine landslide is capable of producing a tsunami that matches the historical observations for the 1852 Banda Sea event, and if such a source is possible, what the corresponding landslide would need to look like.

\subsection{Background}
As an extension to the work of \cite{ringer2021} and \cite{paskett2023}, we focus on submarine landslides as the second leading cause for tsunamis \citep{hamzah2000tsunami}. Submarine landslides were not fully recognized as a possible impetus for destructive tsunamis until 1998. However, the prevailing wisdom shifted after the 1998 Papua New Guinea event provided clear evidence of a landslide-generated tsunami \citep{kawata1999tsunami}. After this shift, researchers reclassified many tsunamis as landslide generated events. These revisited events included the 1929 Grand Banks tsunami \citep{heezen1952turbidity}, the 1979 Nice tsunami \citep{assier2000numerical}, and the 8150 BP Storegga landslide tsunami \citep{bugge1988storegga}, among others. Since several past historical records were compiled prior to the recognition of submarine landslides as possible triggers, previous investigations into tsunami sources still risk potential bias \citep{harbitz2014submarine, lovholt2015characteristics}.

Submarine landslides come in many shapes and forms. They can start on slopes as shallow as one degree and reach velocities as high as 150 $m/s$ (540 $km/hr$) \citep{ward2002suboceanic}.
These landslides are capable of moving large portions of the sea floor hundreds of kilometers. They include events as extreme as the Storegga landslide off the coast of Norway, which had a total volume of 5,600 $km^3$.
This particular slide generated tsunami waves reaching 29 $km$ inland on several islands in the North Atlantic \citep{bugge1988storegga}. The potential for these landslides as well as other non-seismic sources, to generate devastating tsunamis has been a recent focus of active research, particularly in Eastern Indonesia (see \cite{pranantyo2021complex} for instance).  In particular, \citep{brackenridge2020indonesian,nugrahalateral} identify potential scarps in the Makassar Strait of sufficient size to have generated significant tsunamis directly affecting southern and central Sulawesi as well as eastern Kalimantan.  \cite{heidarzadeh2022potential} investigate a recent earthquake and tsunami off of Seram Island which had an uncharacteristically localized large amplitude wave, and they hypothesize a submarine landslide as the actual tsunamigenic source.  

The proximity of the Weber Deep to populated islands in eastern Indonesia further motivates the investigation into submarine landslide-triggered tsunamis in the region.
With a maximum depth of 7.2 $km$, this forearc region is the deepest point of the ocean in the world, excluding trenches.
The prevailing explanation for this extreme geography is a detachment fault called the Banda Detachment \citep{pownall2016rolling}. This slab-rollback feature created a gap of over 60,000 $km^2$ in the ocean floor \citep{pownall2016rolling,Cummins2020}. The myriad slump scars interpreted from sonar images along the Banda Detachment also imply that frequent and recent landslides occur in this region (see \cite{watkinson2017fault} for instance).

We investigate a tsunami observed throughout the Banda Sea in November of 1852 that may have originated from a landslide in the Weber Deep \citep{Cummins2020}. This event occurred before modern instrumental records, so seismographic data cannot verify the magnitude or location of the potential tsunami source. 
However, understanding historical incidents such as this one is crucial to natural hazard risk assessment because the relevant temporal scales for seismic events last hundreds or even thousands of years \citep{ringer2021}. For context, \citep{ringer2021} focuses on this same event but assumes a mega-thrust earthquake caused the 1852 tsunami. Historical observations of severe shaking in the region serve to reinforce this hypothesis, see \citep{fisher2016reconstruction}). Using a Bayesian formulation, \citep{ringer2021} produces a posterior distribution that describes the likely causal earthquake with magnitude near $8.8$ Mw and centroid location in the northeast portion of the Banda Arc. While \citep{ringer2021} found a viable earthquake that matched the observational data somewhat well, \citep{Cummins2020} hypothesized that a submarine landslide in the Weber Deep was the primary source of the 1852 tsunami. We expand upon these results to thoroughly examine and test the hypothesis that a landslide-generated tsunami may best match the recorded observations.

%Their results indicate that the Walanae fault is the most likely source of the tsunami, even over the plausible alternative along the Flores thrust. However, the closest matching tsunami simulation does not fit as well as anticipated with the observations. In particular, the wave heights reported from Bulukumba are significantly larger than any of the simulations performed for either type of earthquake source.

% I moved this paragraph from the forward model to here in the introduction
%A local landslide near Bulukumba, in addition to the earthquake, may account for the discrepancy between the simulated wave heights and observations. This solution is possible because landslide generated tsunamis are typically more localized, which would explain the disparity between the wave heights at Bulukumba and all other locations.
%In order to simulate an event like this one, we must first develop a method similar to the submarine earthquake simulation that is capable of examining submarine landslides.

%We follow a novel approach presented by Ringer, et al.’s \cite{ringer2021} group that applied Bayesian inference to reconstruct the source earthquake parameters based on anecdotal accounts for this tsunami. While this research found a viable earthquake that matched the observational data somewhat well, a similar investigation by Cummins, et al. \cite{Cummins2020} hypothesized that a landslide was the true cause of the same event. We expand upon these results to thoroughly examine and test the hypothesis of a landslide-generated tsunami.

\section{Data}\label{sec:data}
We use a Bayesian approach to sample landslides from a distribution of possible landslide parameter configurations to determine how well the resulting simulated tsunami matches observed reports of the event. We construct observational probability distributions (a key component of the likelihood) of the tsunami based on historical records following \citep{ringer2021}. We also create a distribution of possible landslide parameter configurations, or more simply, the prior distribution, based on data from documented tsunami-generating submarine landslides.

\subsection{Historical Data}\label{ssec:histdata}

\begin{figure}[h]%
\centering
\includegraphics[width=0.9\textwidth]{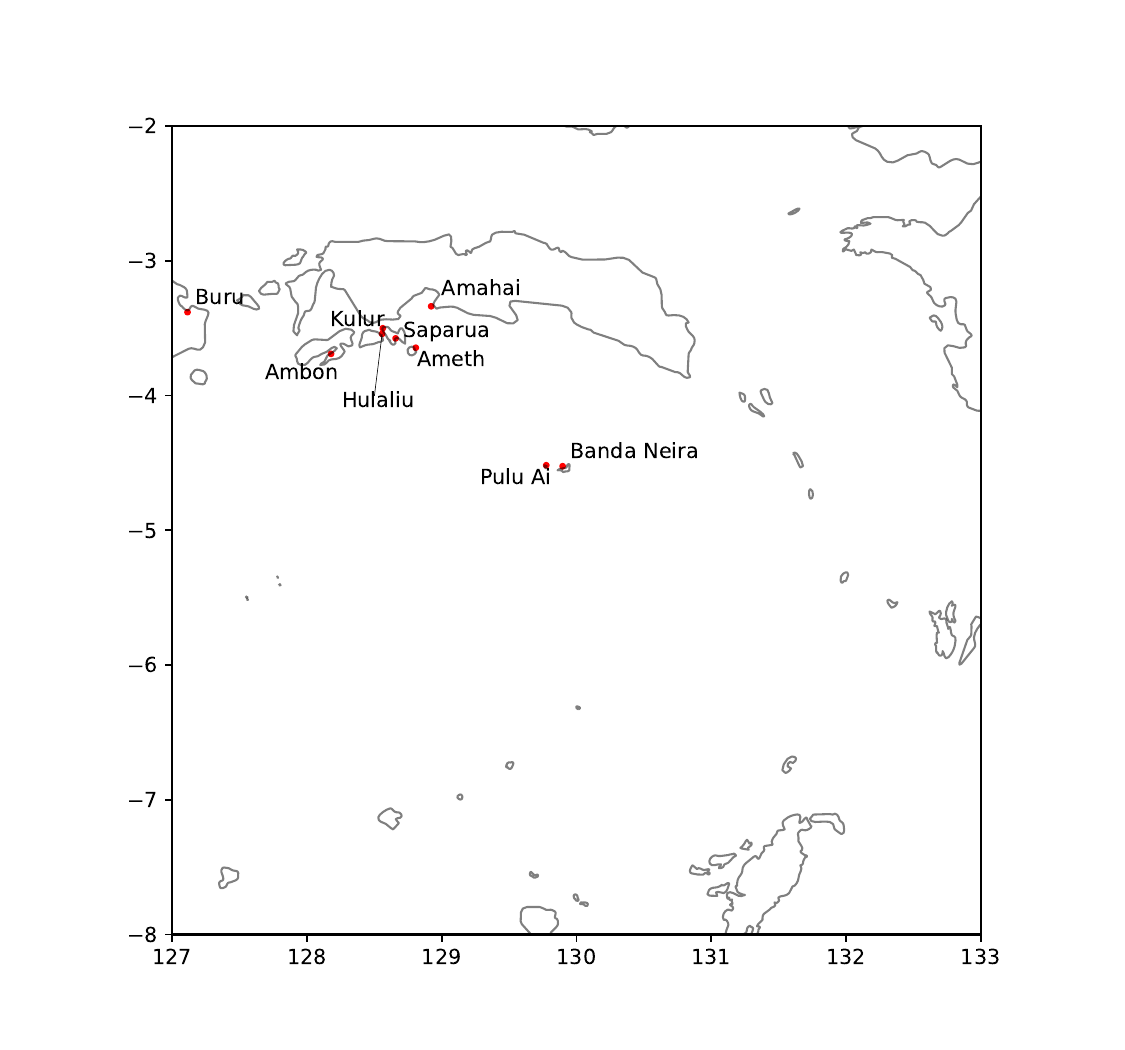}
\caption{The nine observation locations in the Banda Sea for the 1852 tsunami.}\label{fig:obslocs}
\end{figure}

Our simulations rely on anecdotal observations from the 1852 Banda Arc tsunami recorded in accounts from the Wichmann catalog \citep{wichmann1918earthquakes,wichmann1922earthquakes}, which is a compilation of anecdotal historical records from the Dutch colonial era of the Indonesian archipelago translated into English and published recently by \citep{harris2017waves} (2016).
Thirteen observations contain sufficient detail to provide quantifiable information on the tsunami as documented in \citep{ringer2021}.
These accounts span nine locations as shown in Fig. \ref{fig:obslocs}.

Observations from the catalog provide information on:
\begin{itemize}
  \item \textbf{Arrival time}. The arrival time of the first nontrivial wave observed at the specified location.
  \item \textbf{Wave height}. The maximum wave height across all observed tsunami waves from this event. 
  \item \textbf{Inundation length}. The distance the wave traveled on shore.
\end{itemize}
These eye-witness observations are not sufficiently detailed, nor reliable enough to immediately trust the inferred wave heights, arrival times, or inundation lengths.  In essence, we anticipate that the textual observations we make use of, are inherently extremely noisy and uncertain.

\begin{figure}[H]
  \begin{center}
      \begin{tabular}{c}
          \begin{overpic}[width=\textwidth]{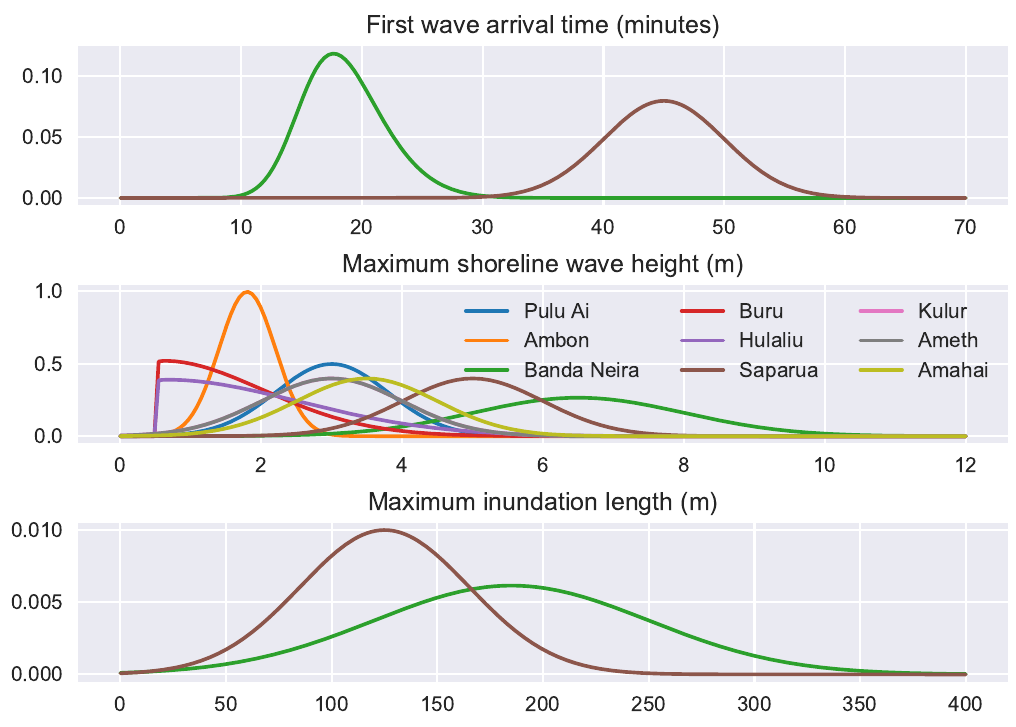}
          \end{overpic}
    \end{tabular}
  \caption{All 13 observational probability distributions. This figure is the same as Fig. 5 in \citep{ringer2021}.}
  \label{fig:likelihood}
  \end{center}
\end{figure}

%# consider combining 'eye witness', changing 'adds' to 'add' and 'in our' to 'to our' response: the plurality of "adds" should refer to "a lack" but if that is unclear we can reword the sentence
%We recognize that a lack of modern measurements and the unreliability inherent in eye-witness accounts adds additional uncertainty to our problem.
%JPW: add a transition here
To address the ambiguity present in such anecdotal accounts, we utilize the probability distributions created in \citep{ringer2021}, which interpret the anecdotal record to assign a probability to each of the observed quantities: arrival time, wave height, and inundation length.
Fig. \ref{fig:likelihood} displays the observational probability distributions for each observation, and is the same as Fig. 5 in \citep{ringer2021}. To illustrate how these distributions are created, we focus on one particular observation location and describe how the corresponding distribution is parameterized.

\subsubsection{Sample Observational Account: Banda Neira}\label{sssec:sampleobs}
Page 242 in the Wichmann catalog provides a brief record of the tsunami at Banda Neira: “Barely had the ground been calm for a quarter of an hour when the flood wave crashed in...The water rose to the roofs of the storehouses and homes...[the wave] reached the base of the hill on which Fort Belgica is built on Banda Neira.”
From this account we find:

\begin{itemize}
  \item \textbf{Arrival time}. ``A quarter of an hour.'' Without a more thorough understanding of the tsunami source, we will assume that the initial earthquake instigated the potentially causal submarine landslide. The historical record indicates that the shaking lasted about 5 minutes, so a landslide could have occurred at any point in this interval.
  Since observations report arrival times after the shaking subsided, we construct our likelihood with a mean of 15 minutes and a skew toward longer arrival times to account for the possible delay. 
  The final distribution is a skew-normal distribution with a mean of 15 minutes, standard deviation of 5 minutes, and skew parameter of 2.
  \item \textbf{Wave height}. ``The roofs of the storehouses and homes.''
  Based on the standard construction of the time, most buildings sat atop stilts with high vaulted roofs. Since this observation occurred on a morning with an exceptionally low tide, we estimate a wave height of about $6.5 m$.
  From this estimation, we consider a normal distribution with a mean of $6.5 m$ with standard deviation of $1.5 m$ to indicate that waves from three to nine meters in height have a reasonable probability. 
  \item \textbf{Inundation length}. ``The water reached the base of the hill on which Fort Belgica is built.''
  To quantify this account, we randomly selected 20 points along the beach and calculated the distance between each point and the base of the specified hill using ARCGIS.
  We used the mean and standard deviation of our measurements to construct the inundation likelihood distribution using a normal distribution with a mean of $185 m$ and a standard deviation of $65 m$.
\end{itemize}

%JPW: transition that states how observations are treated independently
%\subsubsection{Observational Probabilities}\label{subsubsec212}
%By quantifying accounts from the other 8 locations in a similar manner, we obtain a set of resultant distributions that describe the expected tsunami while handling the uncertainty associated with each anecdotal record. We treat the resultant distributions independently since the true dependence of each observation is unknown and difficult to ascertain. This assumption allows us to compute the total likelihood as the product of each observable likelihood.
%\begin{equation}
%  \label{eqn:histlike}
%  P(D|\theta) = \prod_{i=1}^{13} \emph{L}_i(G(\theta)),
%  \end{equation}
%where $G(\theta)$ represents the output of our forward model for sample $\theta$, $L_i$ represents our ith observational distribution, and $D$ represents the total likelihood. For a more in depth discussion of the likelihood, see \cite{ringer2021} by Ringer et al.

\subsection{Prior Distribution Data}\label{ssec:priordata}
Our model requires data from past submarine landslides to inform an educated guess on the most likely landslide parameters as a starting point. Since we are working in the Bayesian context, we need to specify this prior distribution on the parameters to describe the landslide itself. Submarine landslides present a unique challenge for detection due to the limitations of modern measurement equipment; they predominantly occur in regions where active sensing below the ocean's surface is not feasible. This factor distinguishes them from earthquakes and aerial landslides, which available technology can more readily detect.
The data that has been collected for submarine landslides typically requires costly oceanographic surveys \citep{elverhoi2010submarine}, and is commonly restricted to information on the extent of the landslide scarp but not the velocity or precise initial start location of the slide.  It is exceedingly rare to have observational information in real time on submarine landslides which would be necessary to determine the velocity etc.. These limitations mean that the existing data used to parameterize the size of submarine landslides is very limited, and data on the velocity and acceleration of landslides is even more limited. Since only a fraction of submarine landslides are capable of producing an observable tsunami, we have even less data available to quantify what types of submarine landslides are feasible for the current study. We nevertheless apply the data that is available to extrapolate a prior distribution capable of encapsulating the parameters of potentially causal submarine landslides for the tsunami event in question.

We rely on the available scientific literature to construct the prior distribution on the submarine landslide thought to generate the 1852 Banda Sea tsunami. We build these distributions as described in Section \ref{ssec:construction} according to twenty-five events from four sources \citep{harbitz2014submarine,lovholt2015characteristics,urgeles2013submarine,gamboa2021database} recorded from locations across the globe. We also rely on numerical simulations performed in similar research to provide additional reference for the appropriate order of magnitude for the parameters that we identified to model the landslide.

\begin{table}[h]
\caption{Measurement of initial slope and total volume of the slide. All of this data is taken from the Mediterranean Sea.}
\label{tab:reformatted_volume}
    \begin{tabular}{@{}lll@{}}
        \toprule
        Slope ($^\circ$) & Volume ($km^3$) & Source \\
        \midrule
        1.21 & 1.5 & \citep{urgeles2013submarine}\\
        1.45 & 1.5 & \citep{urgeles2013submarine}\\
        1.92 & 0.216 & \citep{urgeles2013submarine}\\
        1.03 & 1.261 & \citep{urgeles2013submarine}\\
        0.62 & 7.064 & \citep{urgeles2013submarine}\\
        0.72 & 4.26 & \citep{urgeles2013submarine}\\
        0.91 & 34.386 & \citep{urgeles2013submarine}\\
        0.75 & 7.134 & \citep{urgeles2013submarine}\\
        1.08 & 27.932 & \citep{urgeles2013submarine}\\
        0.72 & 55.571 & \citep{urgeles2013submarine}\\
        2.03 & 0.037 & \citep{urgeles2013submarine}\\
        1.33 & 0.011 & \citep{urgeles2013submarine}\\
        0.28 & 33.009 & \citep{urgeles2013submarine}\\
        0.28 & 16.951 & \citep{urgeles2013submarine}\\
        1.49 & 4.4 & \citep{urgeles2013submarine}\\
        1.53 & 0.718 & \citep{urgeles2013submarine}\\
        3.98 & 20.563 & \citep{urgeles2013submarine}\\
        4.96 & 3.142 & \citep{urgeles2013submarine}\\
        1.19 & 0.005 & \citep{urgeles2013submarine}\\
        2.16 & 20.0 & \citep{urgeles2013submarine}\\
        7.09 & 0.005 & \citep{urgeles2013submarine}\\
        \bottomrule
    \end{tabular}
\end{table}

\begin{table}[h]
\caption{Estimated thickness of submarine landslides in various circumstances.}
\label{tab:reformatted_thickness}
    \begin{tabular}{@{}lll@{}}
        \toprule
        Thickness ($m$) & Source & Location \\
        \midrule
        25 & \citep{urgeles2013submarine} & Mediterranean Sea \\
        35 & \citep{urgeles2013submarine} & Mediterranean Sea \\
        30 & \citep{urgeles2013submarine} & Mediterranean Sea \\
        30 & \citep{urgeles2013submarine} & Mediterranean Sea \\
        75 & \citep{urgeles2013submarine} & Mediterranean Sea \\
        90 & \citep{urgeles2013submarine} & Mediterranean Sea \\
        10 & \citep{urgeles2013submarine} & Mediterranean Sea \\
        100 & \citep{lovholt2015characteristics} & Simulation \\
        130 & \citep{harbitz2014submarine} & Norway \\
        350 & \citep{gamboa2021database} & Iberia \\
    \bottomrule
    \end{tabular}
\end{table}

The review article \citep{urgeles2013submarine} provides a significant summary of submarine landslide events in the Mediterranean Sea. This data functions as the primary basis for our constructed prior distribution as outlined below because it is particularly well formatted and thorough. Moreover, this data also provides a subset of the parameters for landslides that actually caused tsunamis among over one thousand other recorded submarine landslides. 
% from dallin - I think the following is repetitive
In general, submarine landslide data is difficult to find and even more difficult to verify due to the nature of the events in question since submarine environments are much more difficult to monitor than events on land. %Parameters can range in extremes from 0.001 to 20000, and most landslide records contain no more than one or two parameters each. Among all of the values that we found, this data set contained values that were relatively reasonable, but also fairly distributed. We compensated for more unverifiable data with higher variance in the mixture model (see Fig. \ref{fig:thickness}). 

\section{Methods}\label{sec:methods}

The reconstruction of historical seismic events based on anecdotal evidence of tsunami impacts is an inverse problem. Inverse problems are class of problems for determining the parameters of a system based on observed measurements and their effect on the system overall. They arise in a wide variety of areas including medical imaging, remote sensing, signal processing, and geoscience \citep{tarantola2005inverse}. In contrast to an inverse problem, a forward model outputs observable results based on inputs from a known set of parameters and initial conditions. In the context of this study, the corresponding forward problem would be the accurate simulation of the generation and propagation of a tsunami from a parameterized landslide. In this article, we create a simplified forward landslide-induced tsunami model and use that model to infer the geophysical landslide parameters most likely to match both the historical record and our knowledge of submarine landslides in general by approximating a solution to the corresponding inverse problem. Our approach, following \citep{ringer2021,paskett2023} makes use of modern computational techniques in Bayesian inverse problems to reconstruct a probability distribution that represents the most probable landslide parameters.

\subsection{Bayes Theorem and Markov Chain Monte Carlo}\label{ssec:bayes}
We will rely heavily on Bayes' Theorem which can be succinctly stated as:
\begin{equation}
  \label{eqn:bayes}
  p(\theta|d) = \frac{p(d|\theta)p(\theta)}{p(d)}.
\end{equation}
We seek to identify the posterior distribution $p(\theta|d)$, which represents the probability distribution of landslide parameters $\theta$ given the historical observations $d$. We construct this distribution using the likelihood $p(d|\theta)$, which represents the probability that the observed quantities match the historical record given both a particular set of landslide parameters $\theta$, and the prior distribution $p(\theta)$. The prior distribution represents the probability of a given set of landslide parameters occurring without using any knowledge of the historical record to constrain these parameters. 

To evaluate the likelihood $p(d|\theta)$ we construct a forward model that takes a specific set of landslide parameters $\theta$, and simulates the resultant tsunami. We then record the wave heights and arrival times at the same locations where the historical record was observed. Next, we assign probabilities to the simulated observations based on how well they match the historical record using the observational probability distributions depicted in Fig. \ref{fig:likelihood}.  The final value of the likelihood is computed as the product of each of these individual observational probabilities, i.e. we assume that each observation is independently distributed.  This is summarized as:
\begin{equation}\label{eq:likelihood}
    p(d|\theta) = \prod_{i=1}^{13} p_i(G(\theta),d),
\end{equation}
where the $p_i(\cdot, d)$ are the observational probability distributions depicted in Fig. \ref{fig:likelihood} that are created dependent on the observational data $d$, and $G(\theta)$ is the forward model which simulates a tsunami with landslide parameters $\theta$.
We do modify the likelihood from that used in \citep{ringer2021} by incorporating a new forward model of the tsunami propagation generated by submarine landslides rather than relying on the hypothesis that the tsunami is generated directly from seismic uplift or downlift. The $p_i$ distributions are identical to those in \citep{ringer2021} however.  We discuss the modified forward model, which makes up the rest of the likelihood, in detail in Section \ref{ssec:forward}.

We construct the prior distribution with data from published accounts of submarine landslides, which in a sense will provide an initial guess on the potentially physically relevant parameters. In general, we are unable to calculate the computationally expensive denominator in Bayes' Theorem $p(d) = \int p(d|\hat{\theta})p(\hat{\theta}) d \hat{\theta}$, as this value would require integrating across all possible source events. Instead, we seek the relative probability of our parameters $p(\theta|d) \propto p(d|\theta)p(\theta)$.

%In addition, the prior distribution $p(\theta)$ is a set of parameters that describe our best guess about the submarine landslide. We initially base this distribution on expert and empirical knowledge about what is geographically feasible and the source event in question, but the prior gets updated with each iteration to reflect the information gained from the data.

%\subsubsection{Sampling Methods}\label{sssec:sampling}
We determine the posterior distribution of the landslide parameters using sampling methods that generate a sufficient number of samples from the desired distribution to accurately describe it.
In particular, we use a Markov Chain Monte Carlo (MCMC) method that generates a Markov Chain whose stationary distribution is the desired posterior distribution \citep{kaipio2006statistical}. A Markov chain is a randomized model that describes a sequence of discrete events (samples from the probability distribution), where the probability of each event depends only on the most recent event (the previous sample). A proposal kernel (transition probability) governs how the chain moves from one state to the next.

For this investigation, we use a random walk MCMC method with an adjusted Metropolis-Hastings acceptance rule \citep{hastings1970monte, metropolis1953equation}. This approach means that given a current set of landslide parameters $\theta_k$, we propose a new set of parameters $\theta_p$ that are a random normal perturbation away from $\theta_k$. The Metropolis-Hastings step will then accept the proposed set of parameters (allowing $\theta_{k+1} = \theta_p$) with relative acceptance probability $\alpha_{k+1} = \frac{p(\theta_{k+1}|d)}{p(\theta_k|d)}$. If the proposal $\theta_p$ is not accepted, then $\theta_{k+1}=\theta_k$. Critically, the acceptance probability $\alpha_{k+1}$ only depends on the relative posterior probability, and does not require computing the denominator in Equation \eqref{eqn:bayes}.  To finalize the description, we only need to specify the forward model $G(\theta)$ which is incorporated into the likelihood, and the full prior distribution on the landslide parameters.

%To fully sample from the posterior distribution, we need to specify the likelihood and prior distributions. The observational probability distributions in Fig. \ref{fig:likelihood} are the same as those constructed in \cite{ringer2021} from the historical record, and we are using their same interpretation of the recorded events. However, the forward model in the current investigation is fundamentally different, modifying the total likelihood. The new parameter space we focus on (a landslide source instead of an earthquake) also requires specifying a unique prior distribution as we describe below.

\subsection{The Forward Model}\label{ssec:forward}
Several tools already exist for simulating submarine landslides. For instance (\cite{ward2001landslide,ward2002suboceanic,watts2003landslide,grilli2005tsunami,watts2005tsunami,baba2015parallel,yavari2016numerical,wang2021tsunami}, in addition to many other references) introduce a myriad of landslide models that generate tsunamis. For this investigation, however, we require a simplified landslide model that will couple with our existing tsunamibayes (see \cite{zenodov1_1}) code first developed in \citep{ringer2021}. The landslide model needs to adequately capture the physics of a submarine event in order to distinguish between a landslide induced tsunami and a seismic uplift-generated one. It must also run efficiently on a supercomputer in order to facilitate the simulation of thousands of samples from the parameter space. Furthermore, a small number of parameters must be able to uniquely determine the model to allow for reasonable sampling to take place. A low dimensional parameter space will expedite the parameter search to eventually establish a posterior distribution. Not only would a more complicated and realistic model increase the difficulty of our sampling procedure, but the data we use to infer the posterior distribution is not detailed nor precise enough to warrant inferring a highly detailed model. Hence, although there are several robust landslide to tsunami models available as mentioned above, we develop a new, simplified model that couples with the Geoclaw \citep{leveque2011tsunami} software package. We describe the specific parameters we use in the model in Table \ref{tab:parameters}.

%The first step in our submarine landslide simulation is to design a very simplified forward model that models the consequences of a tsunami from a given set of landslide parameters. Although more robust models exist, we cannot justify inferring the parameters of a complicated system when the simulation relies on such uncertain anecdotal data.
%Instead, we seek to balance simplicity with a reasonable adherence to physical principles with the parameters listed in Table \ref{tab:parameters}. Thus the height of the tsunami wave depends only on the dimensions of the landslide body, the initial velocity, and the bathymetry of the region.

\vspace{10pt}
\begin{center}
\begin{threeparttable}[H]
\label{tab:parameters}
    \caption{Model Parameters for the simplified landslide model. We assume the bathymetry values are either constant everywhere, or functions of the local topography of the ocean floor.}
    \begin{tabular}{@{}ll@{}}
        \toprule
        \textbf{Trainable Parameters} & \textbf{Description}\\
        \midrule
            $d$ & Thickness \\
            $V$ & Volume \\
            $v_0$ & Initial Velocity  \\
            $ar$ & Aspect Ratio \\
        \midrule
        \textbf{Bathymetry Values} & \textbf{Description}  \\ 
        \midrule
            $p_s$ & density of dirt  \\ 
            $p_w$ & density of water  \\ 
            g & gravity \\
            $\theta$ & angle of slope \\
            $C_F$ & surface skin friction \\
            $f$ & Coulomb friction \\
        \botrule
    \end{tabular}
\end{threeparttable}
\end{center}
\vspace{20pt}

We split the full forward model into two key parts.
First, we model the seafloor deformation resulting from the slide using the parameters listed in Table \ref{tab:parameters}. We then pass the resultant time-dependent seafloor deformation to the software package GeoClaw \citep{leveque2008high, leveque2011tsunami, gonzalez2011validation, berger2011geoclaw} to propagate the generated tsunami waves. Geoclaw solves the fully nonlinear, two-dimensional shallow water equations via a finite volume discretization with an adaptive temporal and spatial grid. This operation simulates the propagation of the tsunami from its origin to the observational points of interest, as depicted in Fig. \ref{fig:obslocs}.

\begin{figure}[H]
\label{fig:forces}
  \begin{center}
      \begin{tabular}{c}
      \begin{overpic}[width=.9\textwidth]{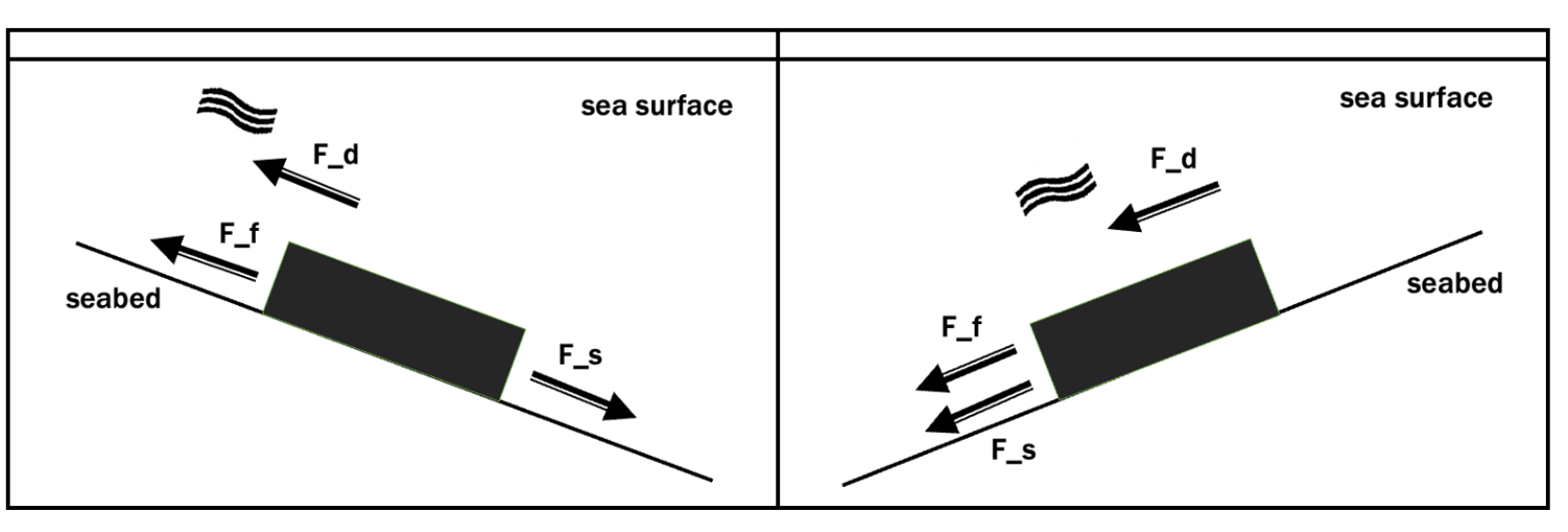}
      \put(15,-2){(a) Forces Upslope}
      \put(65,-2){(b) Forces Downslope}
      \end{overpic}
  \end{tabular}
  \vspace{0.25cm}
  \caption{(a): Depiction of the forces on a block landslide moving uphill. (b): Depiction of the forces on a block landslide moving downhill. }
  \label{fig:2d_slide_movement}
  \end{center}
\end{figure}

%\subsubsection{Seafloor Deformation}\label{sssec:seafloor}
As illustrated in Fig. \ref{fig:2d_slide_movement}, we model the seafloor deformation by modeling the landslide as a solitary block of mass. We derive the motion of this block with respect to the center of mass. Following \cite{pelinovsky1996simplified, lovholt2015characteristics}, we model the landslide center of mass motion via the following equation
\begin{equation}
\label{eqn:slide_motion_down}
    p_sV\frac{\partial u}{\partial t} = (p_s - p_w)gV[\sin(\theta) - f\cos(\theta)] - p_w\frac{C_Fld}{2}u^2.
\end{equation}
Equation \eqref{eqn:slide_motion_down} follows the form $F= ma$ according to Newton's 2nd Law. The expression on the left represents the total force acting on the landslide in terms of mass $p_sV$ and acceleration $\frac{\partial u}{\partial t}$. Three forces determine the velocity of the landslide on the right hand side of the equation as illustrated in Fig. \ref{fig:2d_slide_movement}: 
\begin{itemize}
  \item \( F_s = (p_s - p_w)gV\sin(\theta) \) - Force due to gravity.
  \item \( F_f = (p_s - p_w)gVf\cos(\theta) \) - Force of friction between the seafloor and the landslide block.
  \item \( F_d = p_w\frac{C_Fld}{2}u^2 \) - Drag from the water pushed by the landslide block.
\end{itemize}

Equation \eqref{eqn:slide_motion_down} must also incorporate how the mass will slow down if it starts to move uphill. 
We can represent this change by flipping the sign of the force due to gravity as shown in Fig. \ref{fig:2d_slide_movement}. 
Hence, Equation \eqref{eqn:slide_motion_down} becomes
\begin{equation}
\label{eqn:slide_motion_up}
    p_sV\frac{\partial u}{\partial t} = (p_s - p_w)gV[-\sin(\theta) - f\cos(\theta)] - p_w\frac{C_Fld}{2}u^2,
\end{equation}
when the slope is negative, implying uphill motion.
For simplicity we rewrite Equation \eqref{eqn:slide_motion_down} and Equation \eqref{eqn:slide_motion_up} as
\begin{equation}
    a\frac{\partial u}{\partial t} = b_{\pm} - cu^2,
    \label{eqn:down_simp}
\end{equation}
where
\[
    a = (p_s + p_wV_w), \quad
    b_{\pm}= (p_s - p_w)gV[\pm\sin(\theta) - f\cos(\theta)], \quad\mbox{and}\quad
    c = p_w\frac{C_Flw}{2},
\]
and $b_+$ corresponds to Equation \eqref{eqn:slide_motion_down} and $b_-$ corresponds to Equation \eqref{eqn:slide_motion_up}.

%Similarly, we write \eqref{eqn:slide_motion_up} as
%\begin{equation}
%    a\frac{\partial u}{\partial t} = b - cu^2,
%\label{eqn:up_simp}
%\end{equation}
%where this time
%\[
%    a = (p_s + p_wV_w), \quad
%    b= (p_s - p_w)gV[-\sin(\theta) - f\cos(\theta)], \quad
%    c = p_w\frac{C_Flw}{2}.
%\]

Solving Equation \eqref{eqn:down_simp} is certainly feasible numerically, but it is also possible to write down an exact solution:
\begin{equation}\label{eq:u_soln}
u(t,\theta,v_0) = \left\{\begin{array}{lr}
    
    \frac{-\sqrt{-b}\tan\left(\frac{t\sqrt{-bc}}{a} - \arctan\left(\frac{v_0\sqrt{c}}{\sqrt{-b}}\right)\right)}{\sqrt{c}},
    &  b < 0 \\ \\
    
    \frac{\sqrt{b}\tanh\left(\frac{t\sqrt{bc}}{a} + \operatorname{arctanh}\left(\frac{v_0\sqrt{c}}{\sqrt{b}}\right)\right)}{\sqrt{c}}, 
    & b \geq 0, v_0 < \sqrt{\frac{b}{c}} \\ \\
    
    \frac{\sqrt{b}\coth\left(\frac{t\sqrt{bc}}{a} + \operatorname{arccoth}\left(\frac{v_0\sqrt{c}}{\sqrt{b}}\right)\right)}{\sqrt{c}},
    & b \geq 0, v_0 > \sqrt{\frac{b}{c}}\\ \\
    
    \end{array}\right\},
\end{equation}
where the solution $u(t,\theta,v_0)$ gives us the velocity moving down a slope with angle $\theta$ after $t$ seconds with initial velocity $v_0$. This equation finds the velocity for the center of mass of the slide over the time interval $[0,t]$. We determine the distance traveled over that time interval by numerically integrating $\int_0^t u(t)dt$ using Simpson's rule \citep{humpherys2017foundations}. Note that the discussion thus far has assumed a 2D surface, ignoring the other horizontal direction.

To extend this model so that it applies in both the latitudinal and longitudinal directions, we first need to determine an initial velocity in each direction. This change means that velocity $v_0$ is really vector-valued. 
We can determine how each component is specified by splitting the initial velocity into its latitude and longitude components based on the respective slope gradients. We perform this split by projecting the magnitude of $v_0$ onto the two coordinate axes.
Initial velocity begins the mass's movement in the downhill direction, and the starting point determines the relative slope in the latitude and longitude directions.
After initialization, the model then calculates the distance moved after $t$ seconds by integrating Equation \eqref{eq:u_soln} along each direction, and then recomputes the relative slopes at the new position.
The ending velocity from the previous time step becomes the new initial velocity, and integrating the system once again identifies the next updated position. This process continues iteratively until the landslide center of mass reaches a specified stopping condition as outlined below.

\begin{figure}[H]
\label{fig:3d_slide_movement}
  \begin{center}
      \begin{tabular}{c}
          \begin{overpic}[scale=0.8]{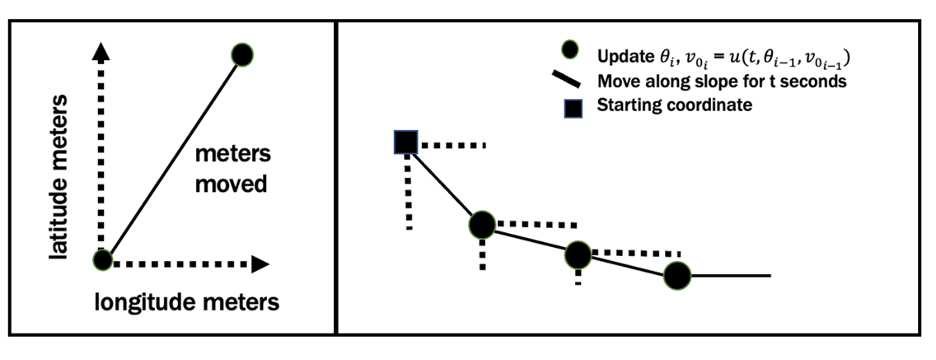}
              \put(2,-1){(a) Distance Moved in One Step}
              \put(45,-1){(b) Distance Moved After Multiple Steps}
          \end{overpic}
      \end{tabular}
      \vspace{0.25cm}
      \caption{(a): Depiction of the total distance (in meters) moved in a single step. This result comes from the calculated distance moved in both longitude and latitude directions. (b): Depiction of the distance moved after multiple steps. The initial latitude/longitude velocity for each step is based on respective final latitude/longitude velocity of the previous step. Additionally, we recalculate the angle of the slope in the latitude and longitude direction at each step.}
  \end{center}
\end{figure}

We include two stopping criteria:
\begin{enumerate} 
    \item \textbf{The velocity falls under a threshold.} We set a default threshold velocity to stop the landslide motion. Thus, if the cutoff velocity exceeds the landslide velocity at any time, we set the velocity to $0$.

    \item \textbf{After ten minutes.} The simulation includes this stopping condition primarily for computational efficiency. However in reality, the initial movement creates the primary and dominant wave, and later motion has a much less significant effect. Thus, the effect of the landslide after ten minutes will have minimal impact on the resultant tsunami wave height.
\end{enumerate}

Once the above procedure determines the path of the landslide's center of mass, we identify a series of discrete points along that path (spaced evenly in the temporal evolution of the slide) and place a `box of mass' with the appropriate dimensions around that point.
For this model, we have the length of the box perpendicular to the direction of motion, and the width parallel to the motion. Geoclaw's default, time-dependent bathymetry routines handle the temporal interpolation between the discrete landslide points.

\begin{figure}[H]
\label{fig:slide_example}
  \begin{center}
      \begin{tabular}{c}
      \begin{overpic}[scale=0.35]{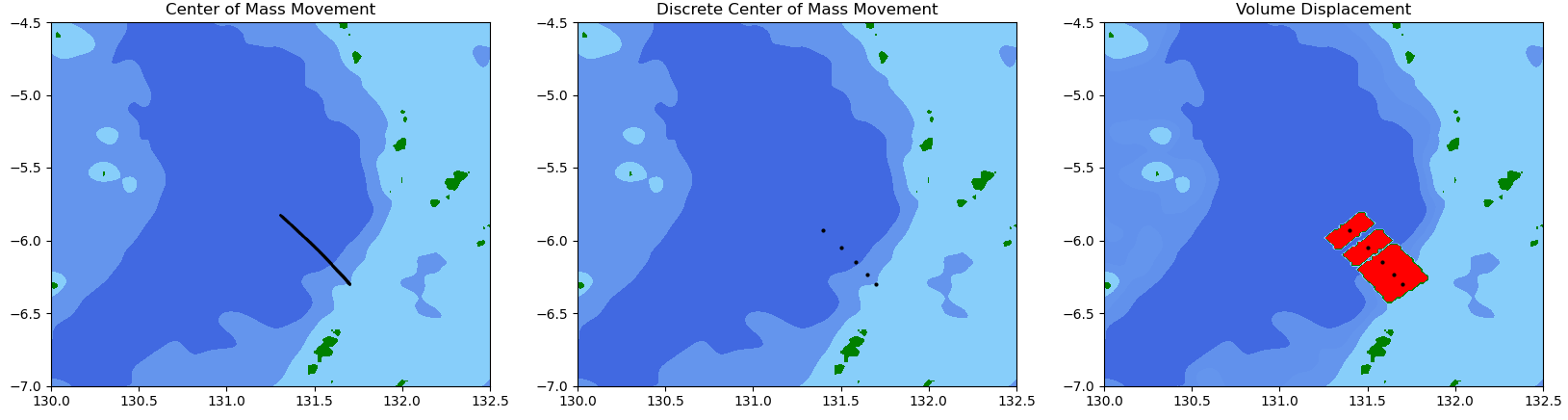}
      \end{overpic}
  \end{tabular}
  \caption{(Left): Depiction of the center of mass movement of the landslide during a ten minute period. (Center): Location of the slide every 120 seconds in five discrete steps. (Right): Volume of the landslide included around the center of mass at five discrete steps.}
  \label{fig:cm_movement}
  \end{center}
\end{figure}

\begin{figure}[H]
    \begin{center}
        \begin{tabular}{c}
        \begin{overpic}[scale=0.35]{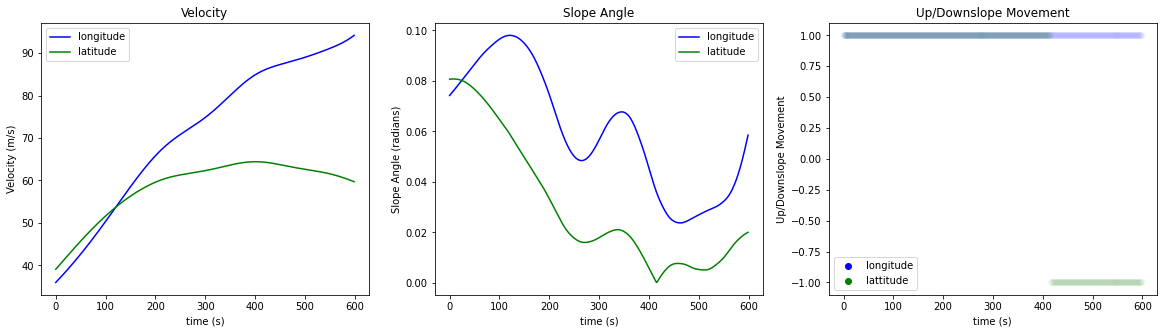}
        \end{overpic}
    \end{tabular}
    \caption{For each graph, longitude and latitude velocity are drawn in blue and green, respectively. 
    (Left): Velocity shown over the initial ten minute period of the slide. 
    (Center): Angle of descent over the initial ten minute period. 
    (Right): Depiction of whether the slide is moving up or down hill. Downhill movement occurs when the value is one, and uphill movement occurs at negative one. The latitudinal velocity begins to slow at four hundred seconds as the slide begins to move uphill.}\label{fig:slide_stats}
    \end{center}
\end{figure}

Fig. \ref{fig:cm_movement} shows the simulation's output for a slide with length, width, and thickness set to $40 km$, $15 km$, and $50 m$, respectively.
The slide has an initial velocity of $25 m/s$ and starts at longitude and latitude values of $(131.7, -5)$. Fig. \ref{fig:slide_stats} visualizes the evolution of the same landslide differently to indicate the temporal evolution of the latitudinal and longitudinal velocities, as well as the absolute value of the slope for each cardinal direction. The final panel in Fig. \ref{fig:slide_stats} indicates when the slide is moving either uphill or downhill. Note that near the end, the slide moves slightly uphill in the latitudinal direction after the angle in the center plot briefly touches zero. At this point the velocity also begins to slow down substantially.

\begin{figure}[H]
  \begin{center}
      \begin{tabular}{c}
      \begin{overpic}[scale=0.5]{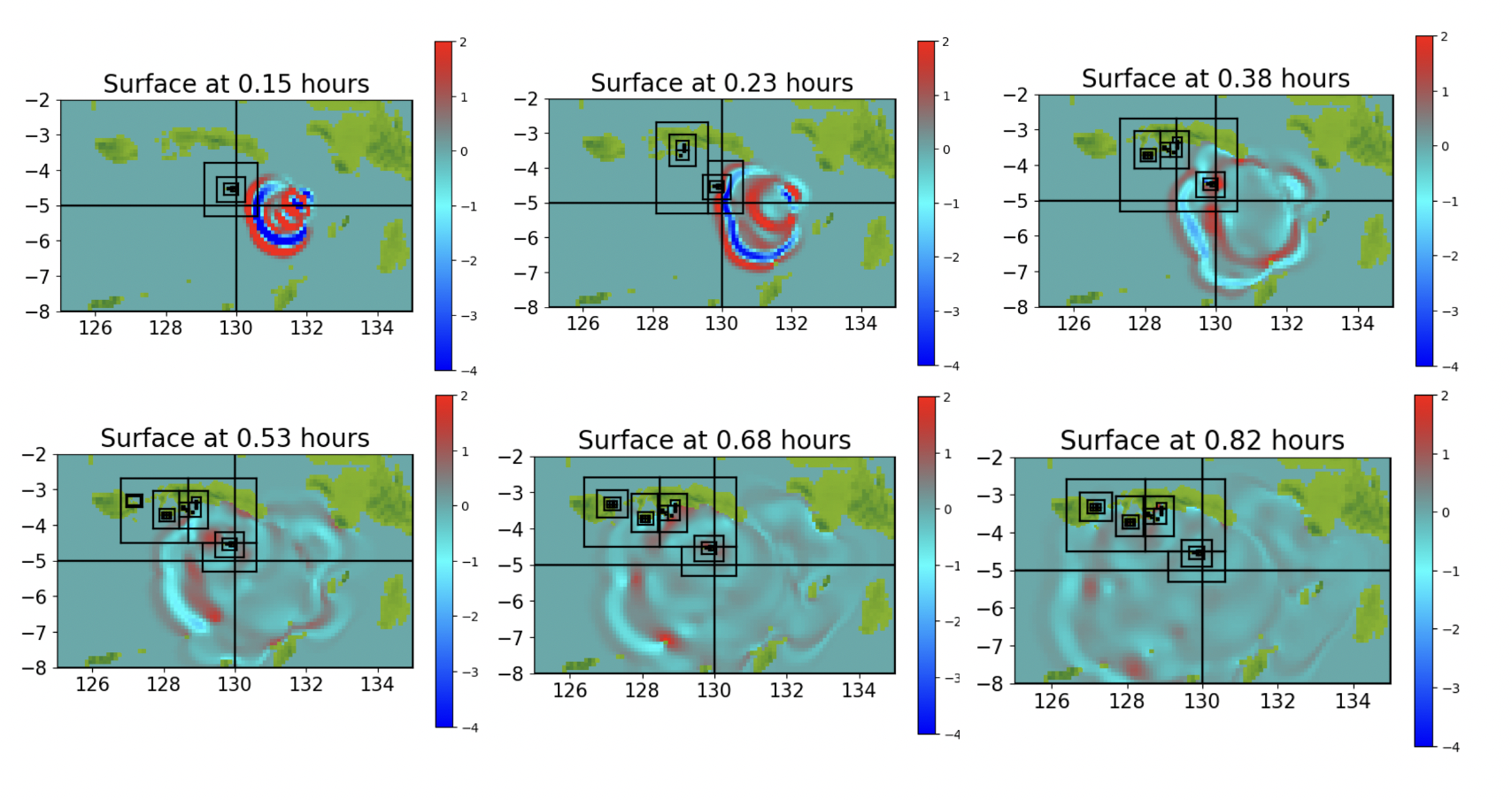}
      \end{overpic}
  \end{tabular}
  \vspace{0.25cm}
  \caption{Tsunami wave propagation generated from the submarine landslide shown in Figs. \ref{fig:cm_movement} and \ref{fig:slide_stats}. The rectangular boxes indicate regions where the adaptive mesh is refined.}
  \label{fig:surface_plot}
  \end{center}
\end{figure}

After calculating the seafloor deformation due to the landslide model, the model passes this modified bathymetry into the Geoclaw software package to propagate the tsunami. In each simulation, we use all the default settings in Geoclaw and employ the same linearized adjoint adaptive mesh \citep{davis2017adjoint} strategy used in \citep{ringer2021,paskett2023}. This mesh includes six different levels of refinement going from the coarsest six arcminute resolution down to three arcseconds at the finest resolution.
Fig. \ref{fig:surface_plot} depicts the simulated waves generated from the landslide shown in Figs. \ref{fig:cm_movement} and \ref{fig:slide_stats} at several different time steps. Most of the wave dissipates within an hour and we can record the maximum wave heights at each observation location. We then insert the recorded wave arrival times, maximum wave heights, and inundation lengths (as discussed in \citep{ringer2021}) into the observational probability distributions depicgted in Fig. \ref{fig:likelihood}. Finally, we define the product of the resultant probabilities as the likelihood probability as described in \eqref{eq:likelihood}.

\subsection{Constructing the Prior Distribution}\label{ssec:construction}
We construct the prior distribution for all of the landslide parameters using the data discussed in Section \ref{ssec:priordata}. The data described there is converted into a probability distribution describing the most probable values of submarine landslide parameters. 
We consider the model parameters as statistically independent, so the prior distribution for a sample $\theta$ is the product of the prior probability for each parameter,
\begin{equation}
  \label{eqn:priorlike}
  P(\theta) = \sum_{j=1}^5 p_j(\theta_j),
  \end{equation}
where, for example, $p_1(\theta_1)$ corresponds to a probability distribution on the total volume of the submarine landslide.

\begin{figure}[H]
  \begin{center}
      \begin{tabular}{c}
      \begin{overpic}[scale=0.8]{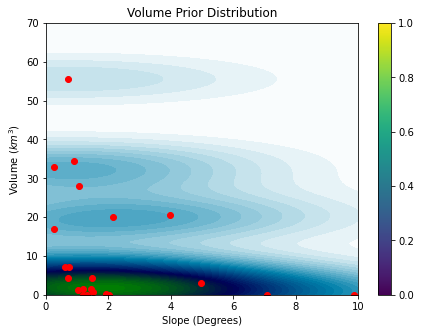}
      \end{overpic}
  \end{tabular}
%  \vspace{0.25cm}
  \caption{Joint prior distribution for the volume and initial slope of a submarine landslide. Red points represent published values of past submarine landslide events pulled from Table \ref{tab:reformatted_volume}. We construct the prior distribution by placing a Gaussian distribution around each of these red points. We also truncate the distribution to allow for only positive values of both the slope and volume. The colorbar on the right indicates the relative prior probability for a given set of parameters.}\label{fig:volume}
  \end{center}
\end{figure}

The volume of the submarine landslide is correlated with the slope of the initial point of movement \citep{ward2001landslide,ward2002suboceanic}. More practically, published records of landslide volumes almost always include slope measurements as well, so combining these two variables together into a single distribution is logical. We use the data shown in Table \ref{tab:reformatted_volume} to construct the prior distribution on volume and slope as a joint Gaussian mixture model, i.e.
\begin{equation}
  \label{eqn:vollike}
  p_{s,v}(\theta_s,\theta_v) = \sum_{j=1}^{21} \emph{N}(\theta_s,\theta_v;\mu=d_j,\Sigma = 4 I_2)\pi_j,
  \end{equation}
where $\theta_s$ refers to the initial slope parameter and $\theta_v$ is the volume of the slide.
In this equation, $N(\cdot,\cdot;\cdot,\cdot)$ indicates a 2D normal (Gaussian) distribution with mean $\mu$ and covariance $\Sigma$ (in this case just 4 times the identity matrix $I_2$). The resultant distribution is a sum of these 2D Gaussian distributions centered on each of the data points given in Table \ref{tab:reformatted_volume}, each with diagonal covariance of $4$. This covariance was a choice we made to ensure that the resultant distribution makes a relatively smooth connection between the existing data points in Table \ref{tab:reformatted_volume}, and the value of $4$ (in degrees for the slope, and $km^3$ for the volume) in both directions is not indicative of any specific restriction, but is simply a modeling choice. 
Without any specific domain knowledge to prefer any subset of the data presented in Table \ref{tab:reformatted_volume}, we set the $\pi_j$ to equally weight each data point to ensure that $p_{s,v}$ is adequately normalized, i.e. $\pi_j = \pi$ is the same for each value of $j$. The resultant joint prior distribution on slope and volume is depicted in Fig. \ref{fig:volume}.

We determine length and width from volume and thickness using an aspect ratio parameter. Without any additional prior knowledge on the statistics of aspect ratios for underwater landslides, we did not rely on any specialized distribution other than a uniform one.
Hence, the prior distribution for the aspect ratio is uniform from $0.3$ to $1$, ensuring that the length of the slide is perpendicular to the direction of movement. We always measure the width along the direction of active motion. This range prevents the width from shrinking unrealistically small with an aspect ratio less than $0.3$.

We use the data in Table \ref{tab:reformatted_thickness} to create a prior distribution on the thickness of submarine landslides using a Gaussian mixture model technique similar to that used for volume and slope angle, but in this case we are restricting our attention to a single variable:
\begin{equation}
  \label{eqn:thicklike}
  P(\theta_i) = \sum_{j=1}^{10} \emph{N}(\theta_i;\mu=d_j,\sigma = 10)\pi_j.
  \end{equation}
Again, $N(\cdot;\cdot,\cdot)$ represents a normal distribution with mean $\mu=d_j$, the value of the jth data point from Table \ref{tab:reformatted_thickness} and variance $\sigma=10$. We chose $\sigma$ sufficiently large to ensure nonzero values between reference data points, and we select $\pi_j$ to place equal emphasis on each distribution so that once again $\pi_j=\pi$ is the same for each value of $j$.

\begin{figure}[H]
\label{fig:thickness}
  \begin{center}
      \begin{tabular}{c}
      \begin{overpic}[scale=0.6]{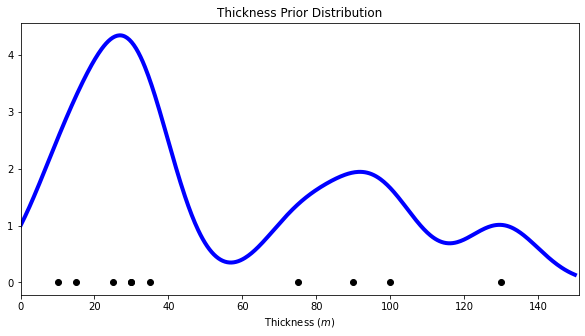}
      \end{overpic}
  \end{tabular}
  \vspace{0.25cm}
  \caption{Prior distribution for the thickness of a submarine landslide. Black points represent values for recorded submarine landslide events as shown in Table \ref{tab:reformatted_thickness}. A Gaussian distribution is placed around each point in the construction of the prior.}
  \end{center}
  \end{figure}

While earthquakes are the most common catalyst for submarine landslides, many other forces can cause unstable slopes to fail.
We do not model the events preceding a submarine landslide, but still incorporate this information in the inference by searching over the slide's initial velocity which is dictated by the initial cause of the slide.
A landslide generated directly from a significant earthquake may have a higher initial velocity than a slide caused by other processes, or a slide that occurs moments after an earthquake destabilizes the slope.
However, there is almost no recorded data on the initial velocity of submarine landslides, so we select a prior distribution for initial velocity with uniform distribution between $10$ and $100$ meters per second.

\begin{figure}[H]
\label{fig:depth}
  \begin{center}
      \begin{tabular}{c}
      \begin{overpic}[scale=0.55]{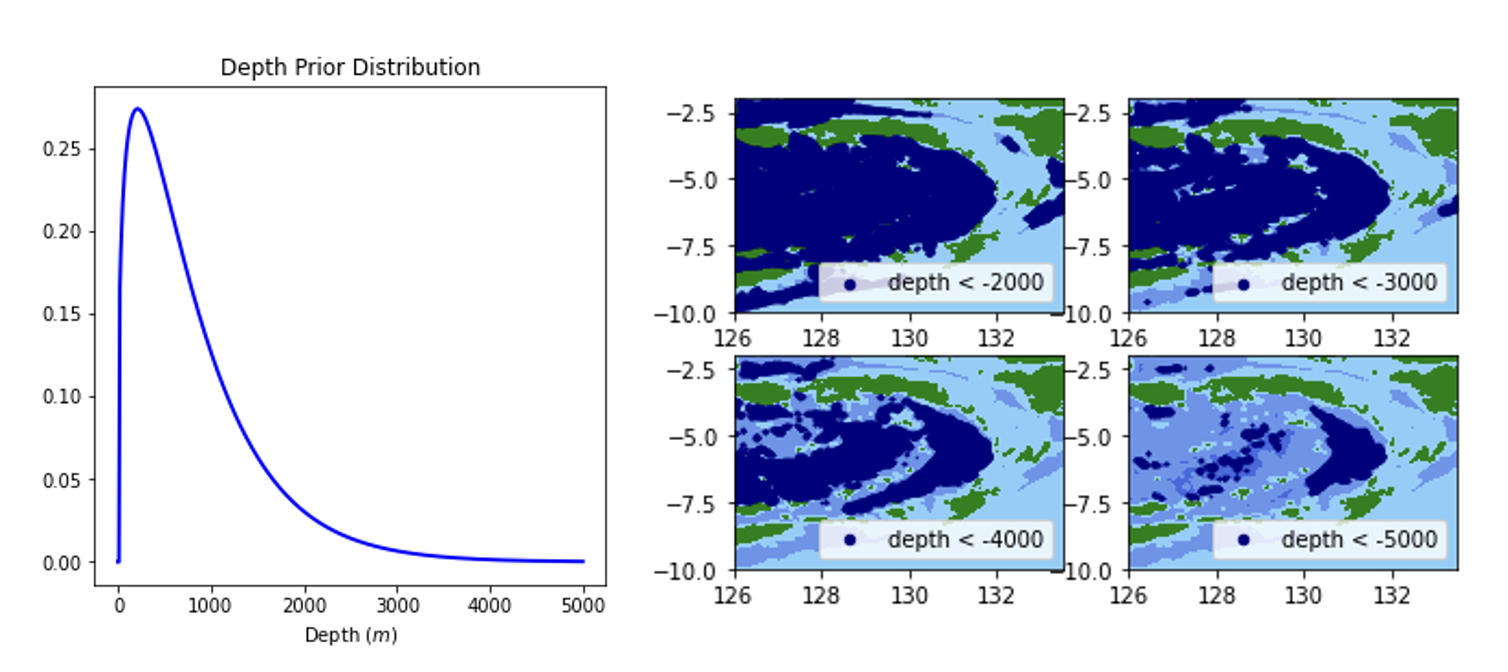}
      \end{overpic}
  \end{tabular}
  \vspace{0.25cm}
  \caption{(Left): Prior distribution for the depth of a submarine landslide. Landslides starting at a shallow depth are more likely to disturb the water's surface and have a higher prior probability. (Right): Latitude/longitude points below a certain depth with prior probability shown in navy.}
  \end{center}
\end{figure}

The prior distribution for latitude and longitude depends on bathymetric depth. Landslides occurring at shallower depths are more likely to cause a tsunami, while landslides deep in the ocean may not create a noticeable disturbance at the surface. To determine a prior distribution for location, we create a chi-squared distribution with Python scipy specific parameters: scale=$300$, df=$2.7$ (degree of freedom), and loc=$250$. This function places the mode at $500$ meters while ensuring a $0$ probability for non-positive depths and a long tail toward deeper values.
Because the depth of the initial position of the slide depends on its geographic location, we create the prior distribution for depth and map its inverse through the bathymetric file to produce the prior distribution for latitude and longitude as depicted in Fig. \ref{fig:depth}.

\subsection{Markov Chain Monte Carlo sampling}\label{ssec:simulation}
We sampled from the parameters' posterior likelihood distributions with ten different Markov Chain Monte Carlo (MCMC) chains running simultaneously on BYU's supercomputer through the Office of Research Computing. Propagating the tsunami waves to the gauge locations with the software package GeoClaw makes up the majority of the computational expense. Each GeoClaw simulation runs in parallel on 24 cores, so a single simulation of the forward model lasts between five and nine minutes of real time computation. The variance in compute times arises from the different computational architectures available through BYU's supercomputing center. The differences in distances depending on the starting latitude and longitude also contributes to this variance. The other required steps in the computation including the accept/reject step and the forward modeling of the submarine landslide compose a small fraction of the computational overhead.

\begin{figure}[H]
\label{fig:start_chains}
    \begin{center}
         \begin{overpic}[width=.75\textwidth]{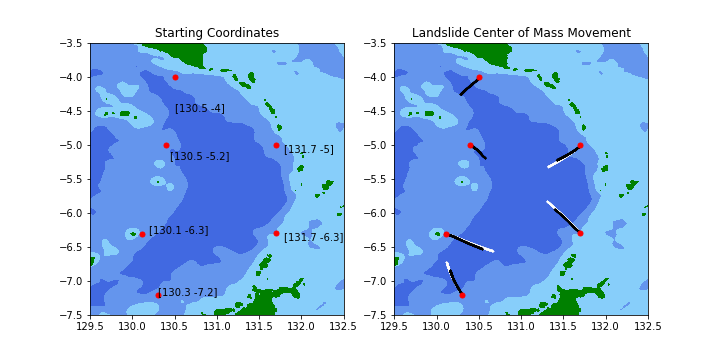}
      \end{overpic}
    \caption{(Left): Starting points for each chain. Chains starting at [130.5,-4] and [130.5,-5.2], close to Banda Neira, begin with reduced volume because they would otherwise produce highly unrealistic tsunamis. (Right): Center of mass landslide movement at initial values for each chain. Black points represent slides with initial velocities of $25 m/s$, and white points represent slides with initial velocity of $75 m/s$.} \label{fig:chains_start}
    \end{center}
\end{figure}

We started the chains at six geographic locations as indicated in Fig. \ref{fig:chains_start}. Table \ref{tab:init_chains} includes all the parameter values for the initialization of each chain. The majority of the chains started with a volume of $30 km^3$ and an aspect ratio of $0.375$ because numerical simulations by Cummins et al. \citep{Cummins2020} found the landslide that best matched observational data was $40 km$ long by $15 km$ wide by $50 m$ thick for a total volume of $30 km^3$. Chains 9 and 10 started with a reduced volume and initial velocity because these initial points were so close to the observation locations in Banda Neira that a higher initial velocity and volume immediately led to implausible tsunami waves, such as a wave height of $40m$ high in Banda Neira. This wave height was also large enough to cause numerical instability by violating the CFL condition in Geoclaw (see \citep{humpherys2017foundations} for a description of the CFL condition), and such large waves are clearly outside the realm of possibility for the constrained data we have available.

\begin{center}
\label{tab:start_params}
    \begin{threeparttable}[H]
        \caption{Starting parameters for all 10 chains.}
        \label{tab:init_chains}
        \begin{tabular}{@{}ccccccc@{}} 
             \toprule
             Chain & Latitude & Longitude & Initial Velocity (m/s)& Volume (km$^3$)& Thickness (m)& Aspect Ratio \\ 
             \midrule
             1 &-6.2 & 130 & 25 & 30 & 50 & 0.375 \\ 
             2 &-6.2 & 130 & 75 & 30 & 50 & 0.375 \\ 
             3 &.-5 & 131.7 & 25 & 30 & 50 & 0.375 \\ 
             4 &.-5 & 131.7 & 75 & 30 & 50 & 0.375 \\ 
             5 &.-6.3 & 131.7 & 25 & 30 & 50 & 0.375 \\ 
             6 &.-6.3 & 131.7 & 75 & 30 & 50 & 0.375 \\ 
             7 & -7.2 & 130.3 & 25 & 30 & 50 & 0.375 \\ 
             8 & -7.2 & 130.3 & 75 & 30 & 50 & 0.375  \\ 
             9 &-4 & 130.5 & 15 & 10 & 50 & 0.375 \\ 
             10 &-5 & 130.4 & 15 & 15 & 50 & 0.375 \\ 
             \botrule
        \end{tabular}
    \end{threeparttable}
\end{center}

\begin{figure}[H]
\label{fig:chain_velocity}
    \begin{center}
        \begin{tabular}{c}
        \begin{overpic}[width=\textwidth]{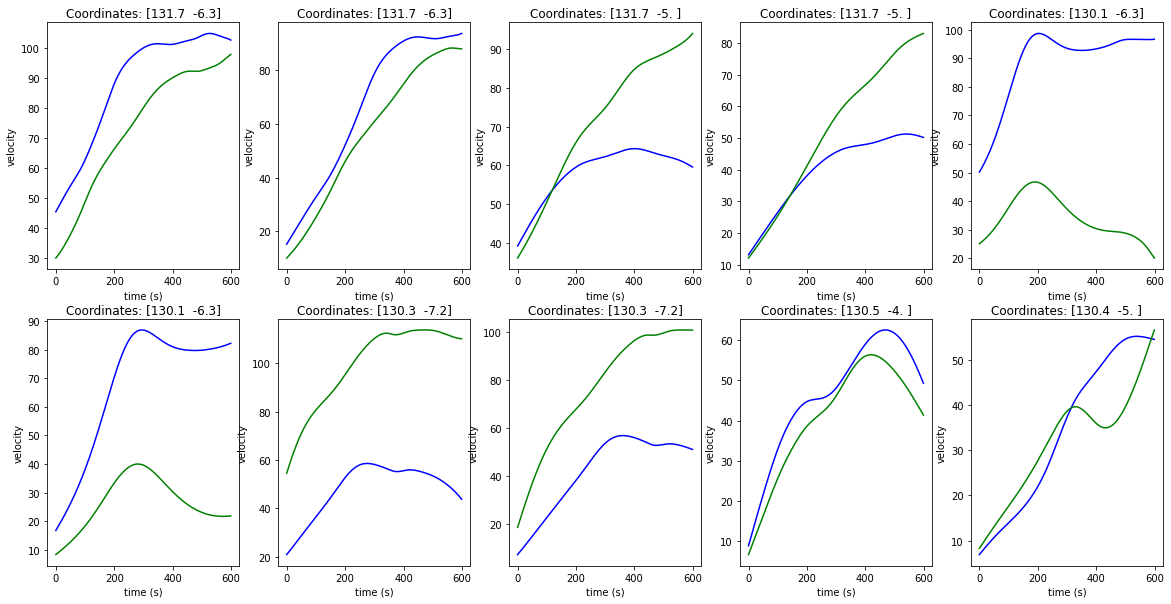}
        \end{overpic}
    \end{tabular}
    \caption{Temporal evolution of the absolute value of landslide velocities for the initial sample from each MCMC chain. The blue curve represents the latitudinal velocity, and green the longitudinal velocity. The slides' velocities increase as the mass of the landslide moves downhill, then slows as the slide reaches the floor of the basin. As already mentioned, these simulations stop the slides after ten minutes. At this point the velocity will rapidly decrease further, but the impact on the tsunami evolution after this time is nearly negligible. Landslide velocities are highly dependent on friction coefficients so we have selected reasonable values.}
    \label{fig:chains_velocity}
    \end{center}
\end{figure}

Figs. \ref{fig:chains_start} and \ref{fig:chains_velocity} depict the initial landslides that initiate the each chain. The distance each landslide travels depends on both the bathymetry, the slope angle, and the selected friction coefficients as indicated in Equation (\ref{eqn:slide_motion_down}) - the equation for landslide motion. Physical limits determine the range of possible friction coefficients. We defer to \citep{pelinovsky1996simplified} where the friction coefficients were set to $f = 0.005$ and $C_D=0.002$. The rough bathymetry data for estimating the angle of descent of the slide makes our landslide simulate artificially fast landslides. In fact, the slides accelerate fast enough that they slosh back up the other side of the Weber Deep, going uphill for several hundred kilometers. To compensate, we choose $C_D = 0.009$,  which gives a reasonable slide path and velocity evolution as seen in the figures below.

\begin{figure}[H]
    \begin{center}
        \begin{tabular}{c}
        \begin{overpic}[scale=.7]{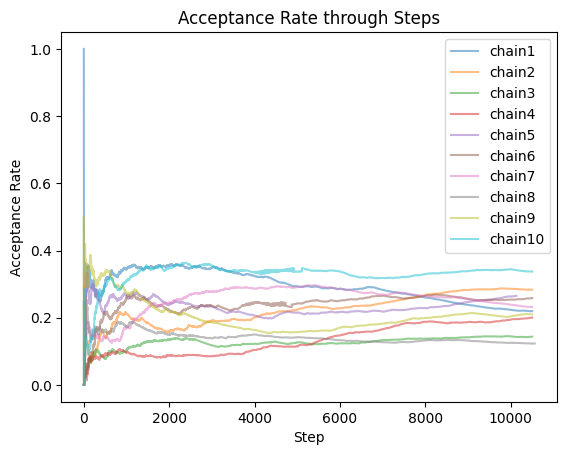}
        \end{overpic}
    \end{tabular}
    \vspace{0.25cm}
    \caption{The acceptance rates after 500 steps vary depending on the chain between 33\% and 12\%. The average across all chains and samples is 21.814\%}\label{fig:kernel}
    \end{center}
\end{figure}

%\subsubsection{Proposal Kernel}\label{sssec:kernel}
As described above, we use the Metropolis-Hasting algorithm to sample from the posterior distribution of landslide parameters. This algorithm determines the movement from one set of landslide parameters $\theta_t$ to another set of parameters at $\theta_{t+1}$ which the algorithm will either accept or reject. We have chosen to work with a random walk proposal kernel given by a Gaussian distributed random variable with mean zero and covariance matrix $\Sigma$: $\theta_{t+1} - \theta_t \sim \emph{N}(0, \Sigma)$. Finding the optimal proposal kernel and the parameter $\Sigma$ is an active area of research. However, conventional wisdom indicates that proposal kernels with acceptance rates of approximately $0.234$ are preferable for random walk MCMC \citep{gelman1997weak}. After experimenting with several different kernels (covariance matrices $\Sigma$), we finally settled on the following values for the entries of the covariance matrix, as these values provided adequate mixing of each chain and led to an acceptance ratio between $0.1$ and $0.35$ over a sufficiently long range of samples (see Fig. \ref{fig:kernel}).

\vspace{0.5cm}
\begin{center}
\label{tab:kernel}
    \begin{tabular}{@{}lllllll@{}} 
         \toprule
          & Lat. & Lon.& Initial Velocity & Volume & Thickness & Aspect Ratio \\ 
         \midrule
         Standard Dev. & 0.075 & 0.075 & 0.95 & 600000000 & 0.5 & 0.001 \\ 
         \botrule
    \end{tabular}
\end{center}

\vspace{0.25cm}

Initially, most of the ten chains did not match the data very well due to the artificial selection of the starting points for the landslide parameters. Even after running several hundred samples for each chain, a significant disparity between the realized likelihood for each chain remained. It was immediately apparent that the less viable chains on the southwest were not going to migrate to the more viable geographic locations to the north. To avoid an overly costly burn-in to reach this type of equilibrium, we ran each of the 10 chains for $2,000$ samples, then resampled each chain with replacement according to posterior probability to re-initialize each chain at a more likely location. This modification concentrated all of our sampling into one of two locations, the western side of the Weber Deep just southeast of Banda Neira, or almost directly east on the other side of the Weber Deep.

\section{Results}\label{sec:results}

\begin{figure}[H]
\label{fig:logpdf_chains}
    \begin{center}
        \begin{tabular}{l}
        \begin{overpic}[width=.8\textwidth]{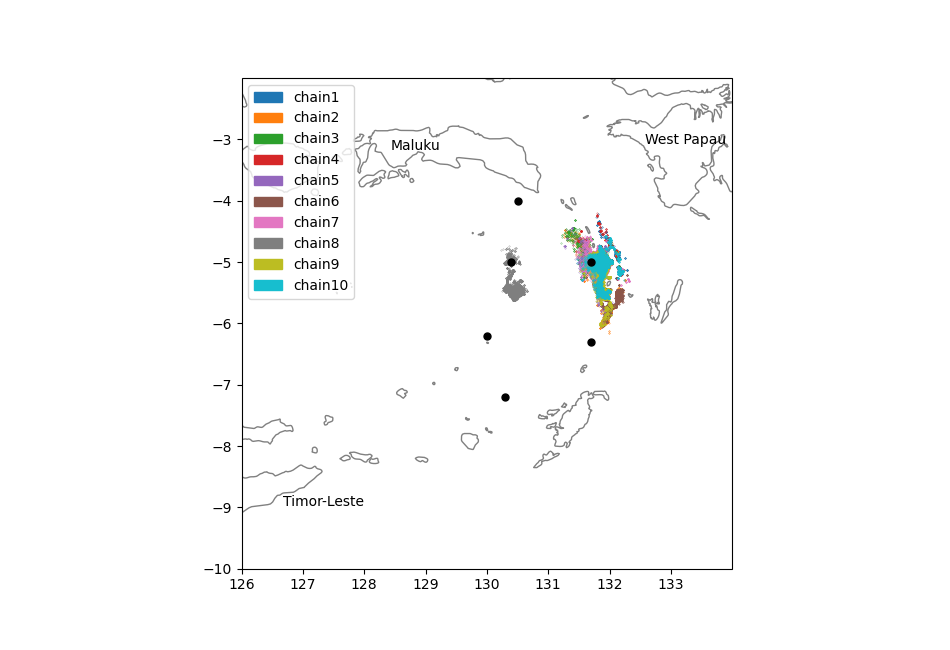}
        \end{overpic}
    \end{tabular}
    \caption{{Latitude-longitude values of the landslide origination colored by chain. Note that only two of the ten chains were selected on the western side of the Weber Deep after the resampling step.}}
    \label{fig:latlon_by_chain}
    \end{center}
\end{figure}

We collected approximately 105,000 samples across ten different chains after the resampling step described above. Fig. \ref{fig:latlon_by_chain} shows the latitude-longitude values of the initial location of the landslide for each chain. The resampling step led to the two different areas on each side of the Weber Deep, with only two of the ten chains remaining on the western edge. In each of these two separate regions, we found that the relevant chains were adequately mixing as the dispersion of the different colors in Fig. \ref{fig:latlon_by_chain} indicates.

\begin{figure}[H]
\label{fig:logpdf}
    \begin{center}
        \begin{tabular}{l}
        \begin{overpic}[scale=.5]{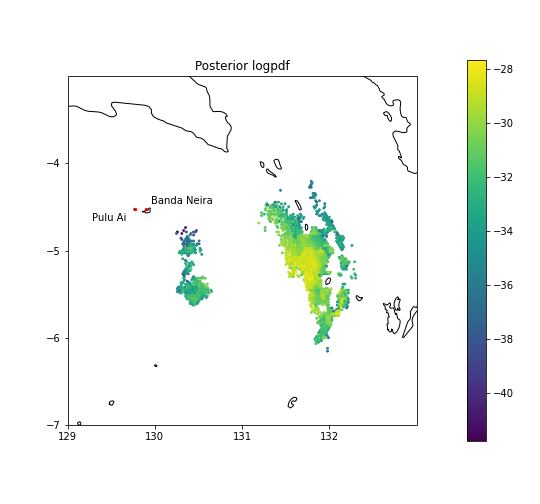}
        \end{overpic}
    \end{tabular}
    \caption{{A zoomed in view of the posterior distribution on latitude and longitude constructed with all of the MCMC samples. Note the areas of high probability on the western and eastern sides of the Weber Deep. The relative probability for a landslide starting on the eastern edge of the Weber Deep is clearly much higher.}}
    \label{fig:logpdf_orig}
    \end{center}
\end{figure}

To better visualize the difference between these two disparate regions in latitude-longitude space, we can visualize the full posterior on landslide start location - weighted by the logarithm of the full (unnormalized) posterior distribution - as depicted in Fig. \ref{fig:logpdf_orig}. A larger value of the posterior and log (base e) posterior indicates a better fit to both the data and prior distribution simultaneously.  Thus, a change in the log posterior from $-28$ to $-40$ as indicated by the two extremes on the colorbar in Fig. \ref{fig:logpdf_orig} equates to an event better matching the data and prior by a factor of over 160,000. In fact, even going from a log probability of $-28$ to $-30$ corresponds to an event that is approximately $7.4$ times less likely. Essentially, although MCMC allows us to explore the full extent of the posterior distribution, the most likely location of the landslide's initial movement will be in the bright yellow region in Fig. \ref{fig:logpdf_orig}.
    
\begin{figure}[H]
    \begin{center}
        \begin{tabular}{c}
        \begin{overpic}[width=\textwidth]{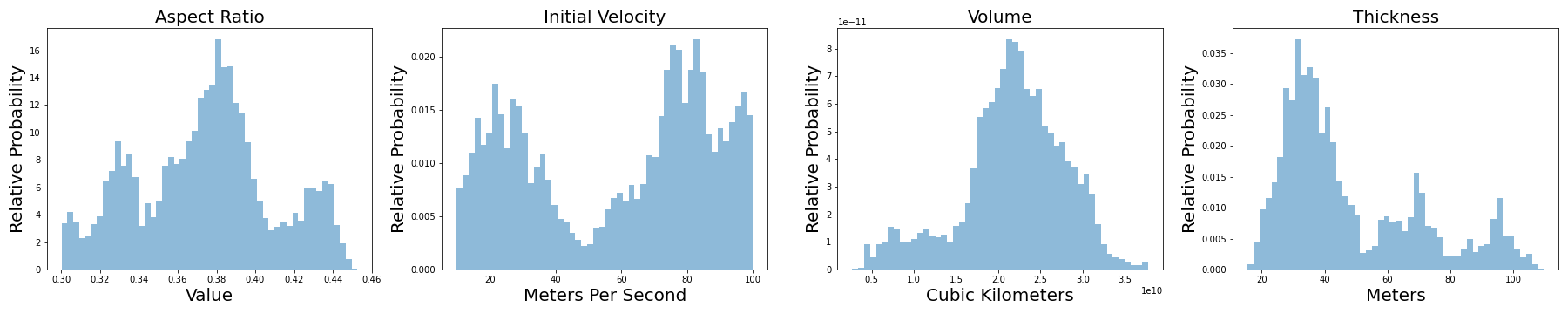}
        \end{overpic}
    \end{tabular}
    \caption{Posterior distribution on landslide parameters constructed by MCMC samples. The vertical axis is a relative sample probability generated in order to normalize the histogram, and these values should not be compared across different parameters.}\label{fig:parameter_post}
    \end{center}
\end{figure}

Fig. \ref{fig:parameter_post} depicts the samples generated from the posterior distribution for the landslide parameters other than latitude-longitude location. Each of these plots is only a one-dimensional slice of the full posterior distribution, and doesn't depict the correlations between the different parameters. The distribution indicates that the posterior has a near universal preferred volume near $2.3 km^3$ , but the other model parameters have far more nuanced posterior distributions. The histograms in Fig. \ref{fig:parameter_post} depict reasonable parameters for a submarine landslide. None of the results that match the observational data (the likelihood) well are significantly unexpected from the scientific literature on landslide-inducing tsunamis.

\begin{figure}[H]
\label{fig:heights}
    \begin{center}
        \begin{tabular}{l}
        \begin{overpic}[scale=.3]{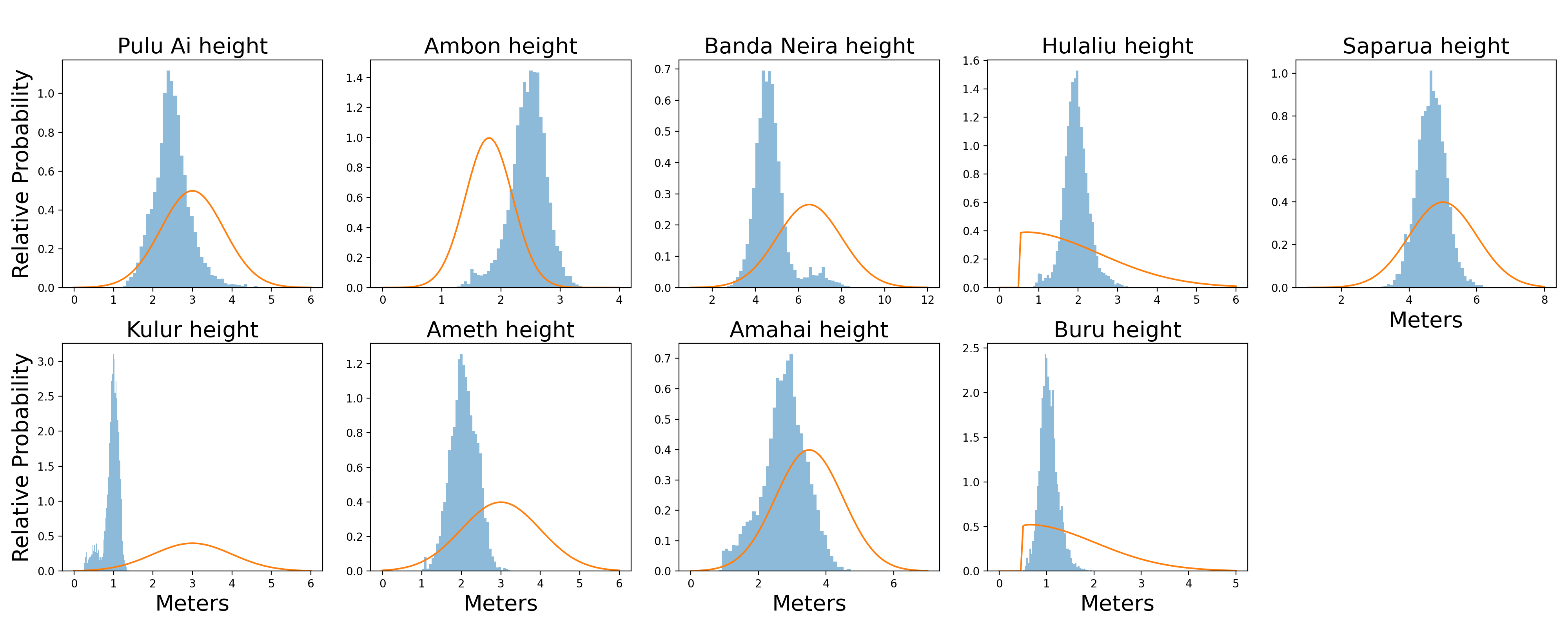}
        \end{overpic}
    \end{tabular}
    \caption{Wave heights generated from tsunami simulations during the MCMC process in blue, compared to observational distributions shown in orange. Note that meters are on the horizontal axis while the relative (non-normalized) probability is on the vertical axis.}
    \label{fig:orig_height}
    \end{center}
\end{figure}

\begin{figure}[H]
\label{fig:arrivals}
    \begin{center}
        \begin{tabular}{c}
        \begin{overpic}[scale=.3]{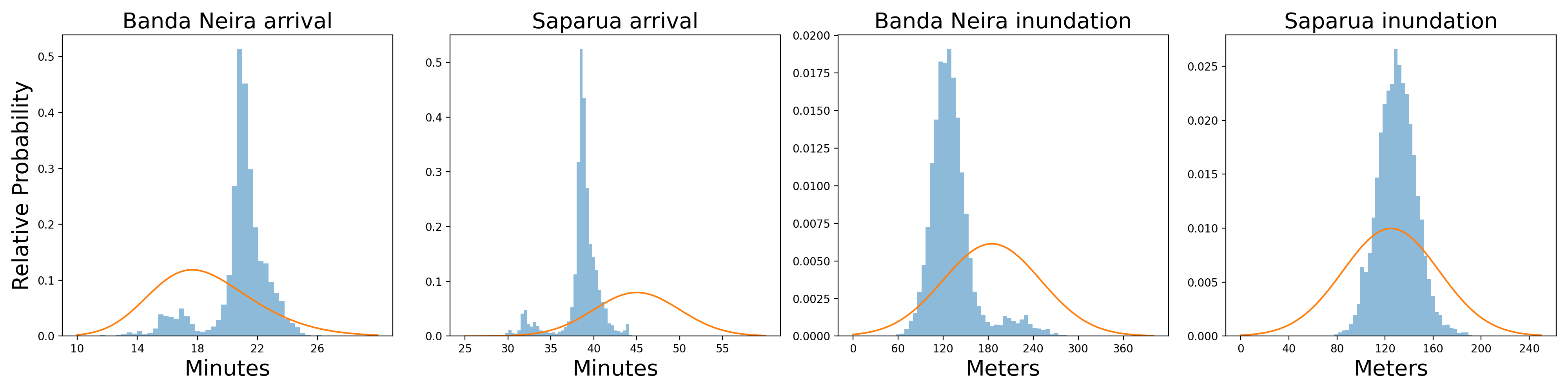}
        \end{overpic}
    \end{tabular}
    \caption{Arrival times and inundation levels generated from tsunami simulations during the MCMC process in blue, compared to observational distributions shown in orange.}
    \label{fig:orig_arrival}
    \end{center}
\end{figure}

In Figs. \ref{fig:orig_height} and \ref{fig:orig_arrival}, we now consider the posterior predictive of the simulated landslide tsunamis whose model parameters are depicted in Figs. \ref{fig:latlon_by_chain} and \ref{fig:parameter_post}. We construct the histograms in these figures with the data sampled from the MCMC chains, whereas the orange curves describe the observational probability distributions first depicted in Fig. \ref{fig:likelihood}. %This data makes up the likelihood in our sampling procedure described earlier and are the same as those created in \cite{ringer2021}. 
%Dallin added after editing:
Note that the vertical axis describes the relative probability while the horizontal axis shows the actual value under consideration. For example, the first panel on the top right of Fig. \ref{fig:orig_height} indicates that the landslide tsunami model is most confident about a tsunami with a $2.5 m$ wave at Pulu Ai while the observational data suggested a $3 m$ wave at Pulu Ai is most likely. In addition, we can see that the standard deviation in the model's output histogram is smaller than that of the observational distribution.

\section{Discussion}
We first note that the clearly preferred initial location of the landslide lies near $-5.2^\circ$ latitude and $131.5^\circ$ longitude as Fig. \ref{fig:logpdf_orig} indicates. This is consistent with the scarp identified by \cite{pownall2016rolling} and further explored in \cite{Cummins2020}, although the preferred initial start of the landslide provided by the statistical sampling here, is slightly to the south and east of the location of that same scarp.  Strict interpretation of the latitude-longitude values of the posterior generated here should be avoided as the model introduced above is not precise enough to determine a precise location of the landslide origin, i.e. the data presented in Fig. \ref{fig:logpdf_orig} matches the observations of \cite{pownall2016rolling,Cummins2020} as well as we would expect.

The landslides starting on the western edge of the Weber Deep do not match the data nearly as well as those on the eastern edge. We can clarify this difference by noting that the posterior predictive, displayed in Figs. \ref{fig:orig_height} and \ref{fig:orig_arrival}, indicates two different modes in the distribution, particularly for wave arrival times. The samples drawn from the eastern edge of the Weber Deep construct the more concentrated mode with a longer arrival time, while the lower mode derives from the two chains that remained on the western edge. While these samples seem to match the arrival time in Banda Neira better, the tsunami arrives far too fast to match the historical record in Saparua. The effect that these two geographic regions have on the maximal wave height at each observation location isn't as immediately apparent, but it appears that the chains on the western edge of the Weber Deep lead to wave heights exceeding the proposed range.

When viewing the posterior predictive, displayed in Figs. \ref{fig:orig_height} and \ref{fig:orig_arrival}, most of the observational distributions appear to have reasonable agreement with the sampled posterior histograms. However, wave heights from the MCMC simulations measured at the island of Kulur do not match the observational distributions exactly. This discrepancy does not elicit considerable concern for two key reasons. First, the constructed observational distributions rely on historical accounts, which introduces uncertainties as discussed above. These include the influence of the observation's precise location in the simulation, the potential modeling error in the selected simulation parameters, or even significant changes in the bathymetry and/or topography of the region. Second, the posterior samples should not match observational distributions precisely because they also incorporate information about what is physically feasible through the prior distribution. 
Furthermore, an exact fit could signify that the predictions actually overfit. If the model were to match the observational probability distributions exactly, it would indicate that our landslide-tsunami model was really a complicated fitting of the data.

The goal of this study was to determine if a physically viable submarine landslide could have generated a tsunami that reasonably matched the observations in the historical record for 1852. Both the posterior distribution on the model parameters and the posterior predictive on the simulated results provide sufficient agreement with our expectations to confirm that a landslide may have generated the tsunami in the historical record. There are of course several caveats to this conclusion, the first of which may be the limitations of the forward model we have used. Specifically, tsunamis created by a landslide require dispersive effects, which Geoclaw does not include as implemented here. In addition, our model doesn't account for the full dynamical effects of a submarine landslide \citep{ward2001landslide,wang2021tsunami}. Such limitations were necessary to have a sampling procedure that would be sufficiently efficient.  These limitations likely indicate that the posterior distribution described here on volume and velocity of the landslide may be over-estimates. Even so, these results do indicate that a landslide may have been the source of the 1852 Banda Sea tsunami.

Despite the limitations on the model and data used here, we note that in addition to the latitude-longitude location of the landslide origin agreeing with geological evidence \citep{pownall2016rolling}, the volume, thickness, and aspect ratio are consistent with the hypothesized source simulated via a very different forward model in \cite{Cummins2020}.  Specifically, \cite{Cummins2020} identified a submarine landslide $40$km long and $15$km wide with a depth of $50$m.  This corresponds to a volume of $30$km$^3$ which is on the high end of the posterior distribution in Fig. \ref{fig:parameter_post}, but well within reason.  The thickness and corresponding aspect ration of $0.375$ also fall within the approximated posterior depicted in Fig. \ref{fig:parameter_post}.  This indicates that our simplified forward model employed in this article is sufficient to capture the relevant parameters of the landslide model, providing a posterior distribution that is consistent with previous efforts.

\begin{figure}[H]
\label{fig:resampled_inun}
    \begin{center}
        \begin{tabular}{c}
        \begin{overpic}[scale=.5]{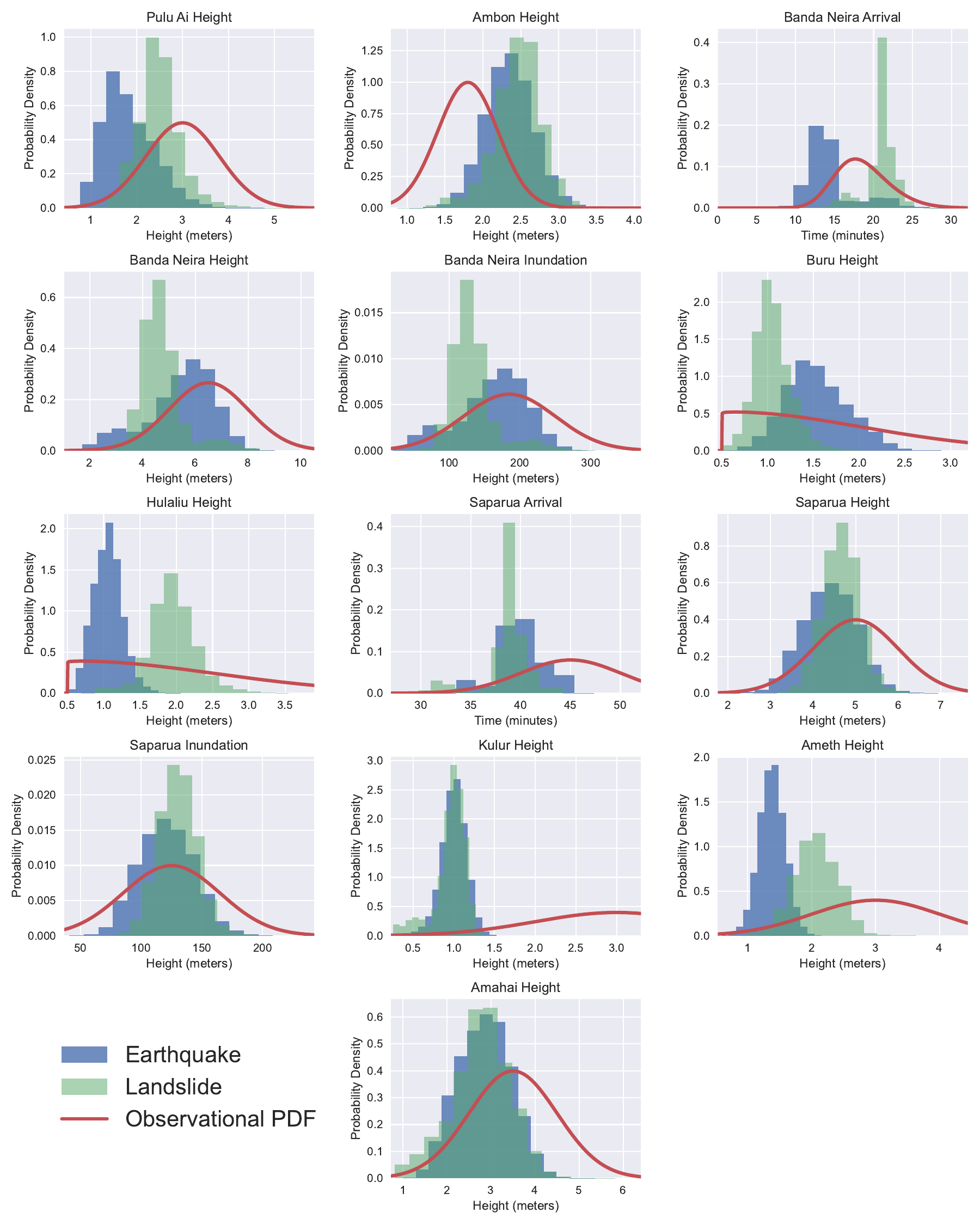}
        \end{overpic}
    \end{tabular}
    \vspace{0.25cm}
    \caption{Observation probability distributions compared to the posterior predictive for earthquake and landslide generated events}
    \label{fig:comparison}
    \end{center}
\end{figure}

None of this discussion quantifies what the cause of the landslide was, nor does it address the multitude of observations of earth shaking in the Banda Sea which are completely neglected here. We can also not directly compare the posterior distribution obtained in \citep{ringer2021} for a hypothesized earthquake source for this 1852 event with the current data because neither posterior distributions (landslide or earthquake generated) are normalized, and so we cannot directly compare the relative probabilities. Since the same observational probability distributions create both posterior distributions, we can visualize the posterior predictive for both in Fig. \ref{fig:comparison} for comparison. Note that some observations are a better match for a landslide source, such as the wave height at Hulaliu. On the other hand, some observations are better fit by an earthquake source, such as the Banda Neira wave height or Saparua arrival time. However, there is no clear `better fit' between the two hypothesized sources.

We trained a Random Forest classifier \citep{pal2005random} on the observations in order to use the classifier to determine whether a particular observation was most likely caused by an earthquake or a landslide. We also compared the distributions directly using KL-divergence. However, both of these methods produced unreliable results, as they would select one hypothesis as a better match solely based on the one observation where the distributions were most distinct such as the wave height in Hulaliu. This behavior is problematic because the components of the posterior predictive that were most distinct were also those observations whose historical record was least reliable. In fact, when these most uncertain individual observations - such as the wave height at Hulaliu - were left out while training the classifier, the classifier flipped its predictions entirely. As a result, it selected an earthquake source 99\% of the time rather than confidently predicting a landslide source as before when all of the observations were included. Therefore, with the current data and available observations, it is not feasible to quantifiably determine whether the source of the 1852 tsunami was a submarine landslide or purely a subduction zone earthquake as suggested in \citep{ringer2021}. However, both options are certainly feasible.

In conclusion, a submarine landslide is a reasonable hypothesis for the source of the 1852 Banda Sea tsunami. According to our analysis and exploration of the potential matching landslides, the most probable landslide had a volume close to $23km^3$, with an aspect ratio around $0.385$ and a location on the eastern edge of the Weber Deep - almost directly east and a little south of Banda Neira. This estimate is not inconsistent with the study of \citep{Cummins2020} where a submarine landslide was first proposed as the source for this event. Further extension of this work would include utilizing the Bayesian technique for a combined earthquake/landslide model. Combined models such as this have successfully matched observational data when a single source model could not \citep{aranguiz20202018}. 
One such tsunami occurred in the northwest portion of the Indian Ocean, where single source models were not able to reproduce the observed near-field run up of 10–12 meters. However, a coupled earthquake-landslide model closely matched the observational data as reported in \citep{heidarzadeh2017combined}. As previously described, the Mw 7.5 earthquake in northwestern Indonesia in 2018 alone does not appear to be the source of the tsunami that devastated the city of Palu.  Multiple submarine landslides are hypothesized to have occurred in Palu Bay which contributed to the destructive power of the tsunami \citep{androsov2023simulating}.

This dual source approach is especially relevant when discussing an earthquake and subsequent tsunami in the Flores Sea in 1820 as described in \citep{paskett2023}. The data generated from the earthquake-based tsunami model does not match the observational distributions as closely as desired for this event. This failure was most apparent when considering the wave heights and arrival times at the observation location near Fort Bulukumba. One way to account for these discrepancies would be to propose a local submarine landslide near Bulukumba, increasing wave heights locally, but not throughout the entire Flores Sea.

A significant weakness of the approach used here and in \citep{ringer2021,paskett2023} is that we have ignored the observational accounts of earth shaking even though there are far more of these accounts than there are of the tsunami in the region. The decision stems from the robustness and widespread understanding of the forward model for tsunamis (Geoclaw), i.e. the mapping from earthquake or landslide parameters to tsunami wave heights is a relatively well-understood problem. In contrast, while there are certainly several models that convert earthquake parameters to shaking intensities, there are none specifically derived for the Banda or Flores Seas, and the general form of such models are rife with uncertainties that would quickly overwhelm the relatively weak signal from the observational distributions. Essentially, the uncertainty in the data is so strong for shaking intensities that we have focused on tsunami observations to eliminate as much uncertainty in the forward model as possible. Even so, inclusion of earthquake shaking into the inversion process is a task that will be undertaken using the forward models that are reasonably available.

\section{Conclusion}\label{sec:conclusion}
We have presented a simplified model of submarine landslide generated tsunamis, and using this model and a Bayesian approach, shown that the 1852 Banda Sea tsunami may have been generated by a submarine slump on the eastern side of the Weber Deep. The simulations presented here are consistent with the literature on submarine landslides, and with the historical, anecdotal evidence, and provide a posterior distribution describing the type of landslide required to best match the data.  Comparing these results with the opposing hypothesis of a source coming from a mega-thrust event along the outer Banda Arc is inconclusive, i.e. either of these hypotheses seem viable from the level of detail provided in the historical record.  The work here does indicate the need to re-examine other historical (and paleo-historic) tsunamis with respect to the potential for a source that is not purely due to seismic uplift.

%% Choose your bibliography style. Some options: plain, unsrt, abbrev etc.
% \bibliographystyle{unsrt}
%% the name of your bib file without the .bib extension. For example, if your file is thesis.bib, you would put \bibliography{thesis}
%\bibliographystyle{sn-mathphys}
%\bibliography{tsunami}


\section*{Statements and Declarations}J.P.W. was partially supported by NSF grant DMS-2206762, and both J.P.W. and R.A.H. acknowledge generous support from Geoscientists Without Borders.  The authors have no relevant financial or non-financial interests to disclose.  R.W. and J.P.W. designed the study, R.W. developed the forward model, and initiated most of the computations with D.S. completing the final computations.  R.W. and D.S. wrote the first draft of the manuscript, and J.P.W. made significant edits.  R.A.H. provided domain expertise and significant guidance throughout the entire process.

%% If you want to include an index, this prints the index at this location. You must 
%% have \makeindex uncommented in the preamble

\end{document}